\pdfoutput=1
\documentclass[review]{elsarticle}
\usepackage{lineno}
\usepackage{hyperref}
\usepackage{amsmath}
\usepackage{graphicx}
\usepackage{xcolor}
\usepackage{subfigure}
\usepackage{rotating}

\modulolinenumbers[5]

\journal{Mathematical Biosciences}

\newcommand{\ds}{\displaystyle}

\begin{document}

\begin{frontmatter}

\title{Practical Identifiability and Uncertainty Quantification of a Pulsatile Cardiovascular Model}

\author[adr1,adr2]{Andrew D. Marquis}
\author[adr2,adr3]{Andrea Arnold}
\author[adr4]{Caron Dean-Bernhoft}
\author[adr1]{Brian E. Carlson}

\author[adr2]{Mette S. Olufsen\corref{mycorrespondingauthor}}
\cortext[mycorrespondingauthor]{Corresponding author}
\ead{msolufse@ncsu.edu}

\address[adr1]{University of Michigan, Ann Arbor, MI}
\address[adr2]{NC State University, Raleigh, NC}
\address[adr3]{Worcester Polytechnic Institute, Worcester, MA}
\address[adr4]{Medical College of Wisconsin, Milwaukee, WI}

\begin{abstract}
Mathematical models are essential tools to study how the cardiovascular system maintains homeostasis. The utility of such models is limited by  the accuracy of their predictions, which can be determined by uncertainty quantification (UQ). A challenge associated with the use of UQ is that many published methods assume that the underlying model is identifiable (e.g. that a one-to-one mapping exists from the parameter space to the model output). In this study we present a novel methodology that is used here to calibrate a lumped-parameter  model to left ventricular pressure and volume time series data sets. Key steps include using (1) literature and available data to determine nominal parameter values; (2) sensitivity analysis and subset selection to determine a set of identifiable parameters; (3) optimization to find a point estimate for identifiable parameters; and (4)  frequentist and Bayesian UQ calculations to assess the predictive capability of the model. Our results show that  it is possible to determine 5 identifiable model parameters that can be estimated to our experimental data from three rats, and that computed UQ intervals capture the measurement and model error.
\end{abstract}


\end{frontmatter}


\section{Introduction}

Precision medicine is a growing model of healthcare that proposes to customize of care, medical decisions, practices, and products to each individual patient. This approach is important, as pathologies such as cancer, autoimmune disorders, and cardiovascular diseases are unique to a given individual making it challenging to develop diagnostic and treatment protocols. One approach, to studying patient-specific complexities,  is to use mathematical modeling to estimate function and predict features that are difficult to measure, thus providing a more comprehensive set of information to distinguish between individual patients.

A rich history of cardiovascular modeling exists in the literature, typically written either from a fluid dynamics perspective (resulting in systems of PDEs)~\cite{FormaggiaCVBook, QuarteroniCVBook, van2011pulse}, to study local flow properties, or from a compartment perspective (resulting in systems ODEs) to study systems level dynamics~\cite{OttesenOlufsenCVBook, ottesen2013development}. This study focuses on analysis of compartment models predicting systems level propagation of flow and pressure through the cardiovascular system~\cite{Blanco10,Kokalari13,Shi11}. In this model type, compartments represent groups of vessels (e.g. large or small arteries or veins,  capillaries, or vessels supplying specific tissues or organs) coupled to a heart compartment that  act as a  pump to drive the system.  Some models include both pulmonary and systemic circulations~\cite{neal2007subject}, while others analyze one of the two systems~\cite{zinemanas1994relating}.  This model can be used to extract vascular properties such as vascular resistance, cardiac contractility, or compliance by fitting models to pressure and flow data from noninvasive imaging studies~\cite{Olufsen05,Pope09,Williams14} and/or from invasive catheterization~\cite{Revie13,pacher2008measurement,mackenzie1979effects} studies.

One of the biggest challenges in calibrating compartment models to data is obtaining accurate parameter estimates. Even in its basic form, where the model is formulated using systems of linear differential equations, forced by a contracting heart, it is typically not possible to uniquely estimate all model parameters. To overcome this, we propose to use sensitivity analysis and subset selection for  {\it a-priori} study of the model structure followed by parameter estimation and uncertainty quantification. In general, parameters that are unidentifiable as  a result of model structure are referred to as \textit{structurally unidentifiable}~\cite{mahdi2014structural}, whereas parameters that are unidentifiable as a result of practical restrictions, such as availability and quality of data, are referred to as \textit{practically unidentifiable}~\cite{miao2011identifiability}. Theoretically, structural identifiability is a prerequisite for practical identifiability. However, in practice it can be difficult to establish the former, since analysis is restricted to  models for which it is possible to define a unique input-output relation~\cite{mahdi2014structural}.

Only a few studies have addressed structural identifiability in cardiovascular models. Kirk et al.~\cite{Kirk13}, studying Windkessel models, showed that three of  four parameters are identifiable, and Pironet et al.~\cite{Pironet16}  demonstrated that every parameter in a linear six-compartment model including a left and right heart, systemic and pulmonary arteries and veins are  structurally identifiable if outputs contain both pressure (in all arteries and veins) and left/right ventricular stroke volume, while models relying on either pressure or volume alone are structurally unidentifiable. Other studies have employed sensitivity  and practical as opposed to structural identifiability analysis in models predicting arterial blood pressure and cardiac output~\cite{Ellwein08,Pope09,Williams14,Gul16}. Several recent studies have addressed uncertainty quantification, mostly for analysis of 1D fluid dynamics models, however to our knowledge, these methodologies have not previously been applied to analysis of compartment models. The study by Eck et al.~\cite{Eck16} develops a guide to uncertainty quantification in cardiovascular models presenting a number of methodologies. Several studies have predicted uncertainties in specific one-dimensional fluid mechanics models~\cite{Cheng13, Arnold16, Eck15, eck2015stochastic, eck2016guide, Paun2018}. Of these, three studies accounted for uncertainty using Kalman filtering~\cite{Cheng13,Arnold16,Eck15}, two used polynomial chaos expansion, and one~\cite{Paun2018} used an MCMC approach based on the Delayed Rejection Adaptation Metropolis (DRAM) algorithm~\cite{haarioDRAM}. To our knowledge, no study has combined these techniques into an organized workflow for the determination of model parameters in compartmental CV models given a specific data set.

In this study, we present a potentially general multi-stage methodology to establish reliable parameter estimation approaches and uncertainty quantification (UQ) for  lumped-parameter cardiovascular models which is used here to characterize left ventricular volume and blood pressure  data from three Sprague Dawley rats. The key steps in our methodology include: (1) the use of literature and available data to compute nominal parameter values specific to each rat; (2) sensitivity analysis and subset selection to determine a set of identifiable parameters; (3) optimization to compute point estimates for the identifiable parameters; and (4)  statistical techniques to quantify uncertainty of the model output.  

\section{Methods}
\label{section:meth}

\subsection{Experimental Data}
\label{Subsection:ExpData}

\begin{figure}[hbt!]
\centering
\includegraphics[width = 10cm]{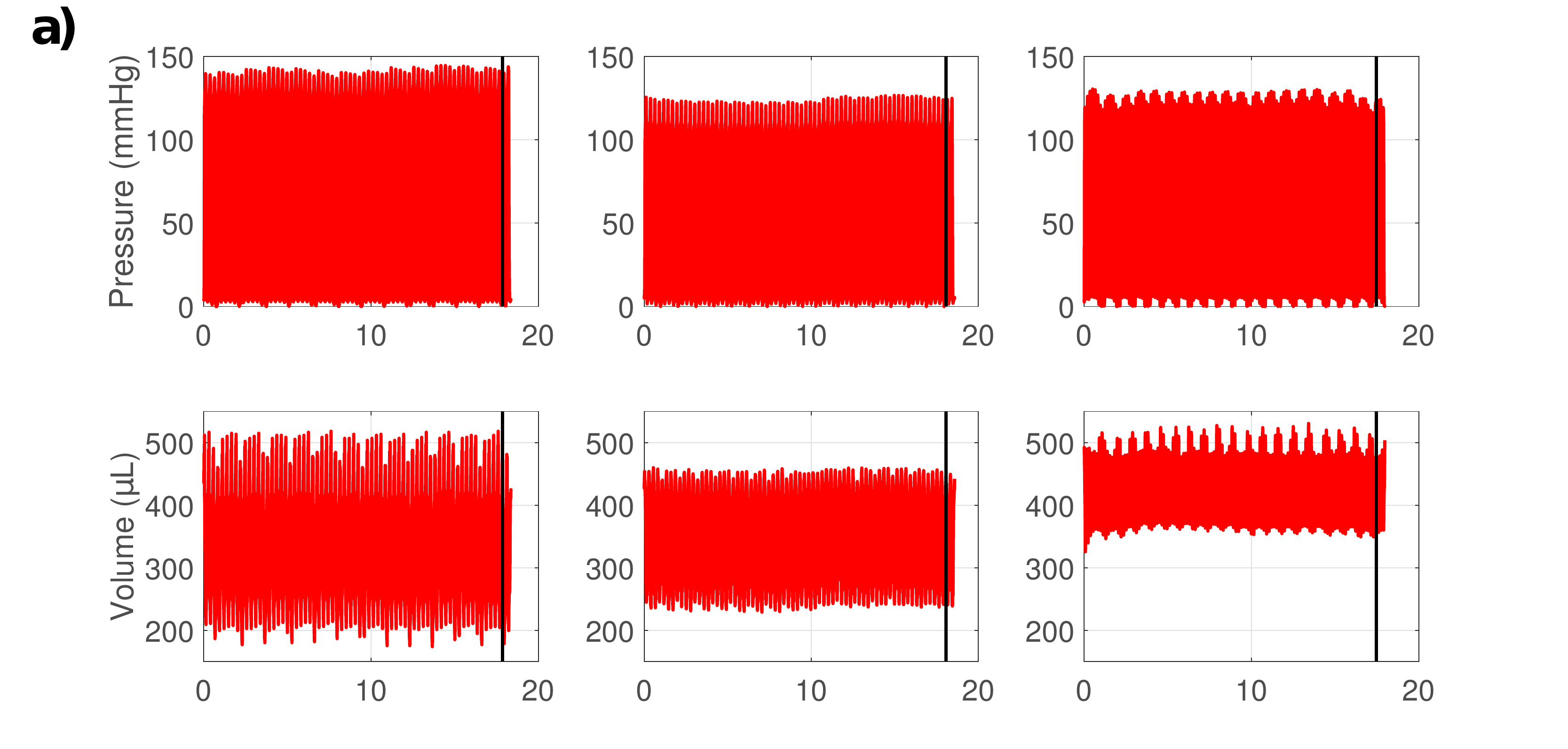}\vspace{-0.35cm}
\includegraphics[width = 10cm]{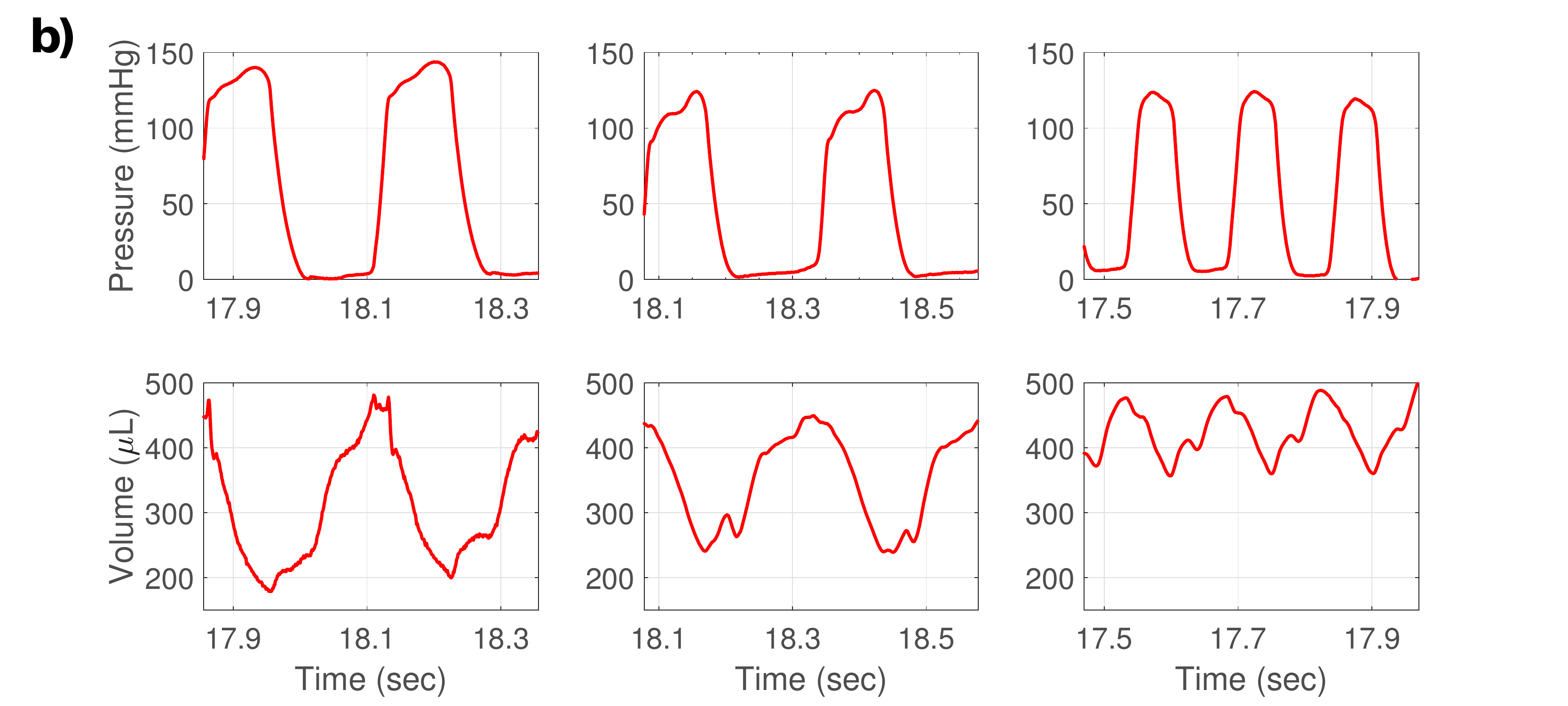}\vspace{-0.4cm}
\caption{Left ventricle pressure and volume hemodynamic data from three rats. Each column corresponds to a different rat. {\bf a)} shows the 20-second raw time-series data  and {\bf b)} shows a zoomed-in view of the final 0.5 seconds used to calibrate the model (marked by vertical black lines on the top two rows.)}
\label{fig:data}
\end{figure}

Data analyzed here are extracted from experiments performed on 3 Sprague-Dawley (SD) rats (2 male, 1 female). The average weight of these animals was $358.0\pm19.6$ g. Rats were anesthetized with sodium pentobarbital (50 mg/kg, ip), and catheters were placed in a femoral vein and artery for administration of anesthetics and monitoring of systemic blood pressure respectively. A pressure-volume conduction catheter (Millar SPR-869, 2F tip with four electrodes and 6mm spacing) was inserted through the right carotid artery into the left ventricle to simultaneously obtain pressure and volume measurements. For each rat basic physiological measures (sex, weight, heart rate, average stroke volume and cardiac output, Table~\ref{tab:data}) were recorded along with continuous measurements of left ventricular volume and pressure.  For this study, approximately 20-second pressure-volume time-series, measured at rest, were selected for model identification and the final 0.5 seconds of each data set was used to calibrate the model, shown in Figure~\ref{fig:data}.
\begin{table}[t!]
\caption{Rat average data.}\label{tab:data}
\centering
\begin{tabular}{cccccc}
\hline
Rat & Sex & Weight & Heart rate & Stroke volume & Cardiac output\\
       & & (g) & (beats/min) & ($\mu$l)  & (ml/min) \\ \hline
Rat 1 & Male & 339 & 240$\pm$3& 308$\pm$1 & 74$\pm$0.2\\
Rat 2 & Male & 350 & 240$\pm$3& 216$\pm$1 & 52$\pm$0.2\\
Rat 3 & Female & 342 & 420$\pm$3& 143$\pm$1 & 60$\pm$0.2\\ \hline
\end{tabular}
\end{table}

\subsection*{Volume-Conductance Calibration}
\begin{figure}[h!]
\centering
\includegraphics[scale=0.3]{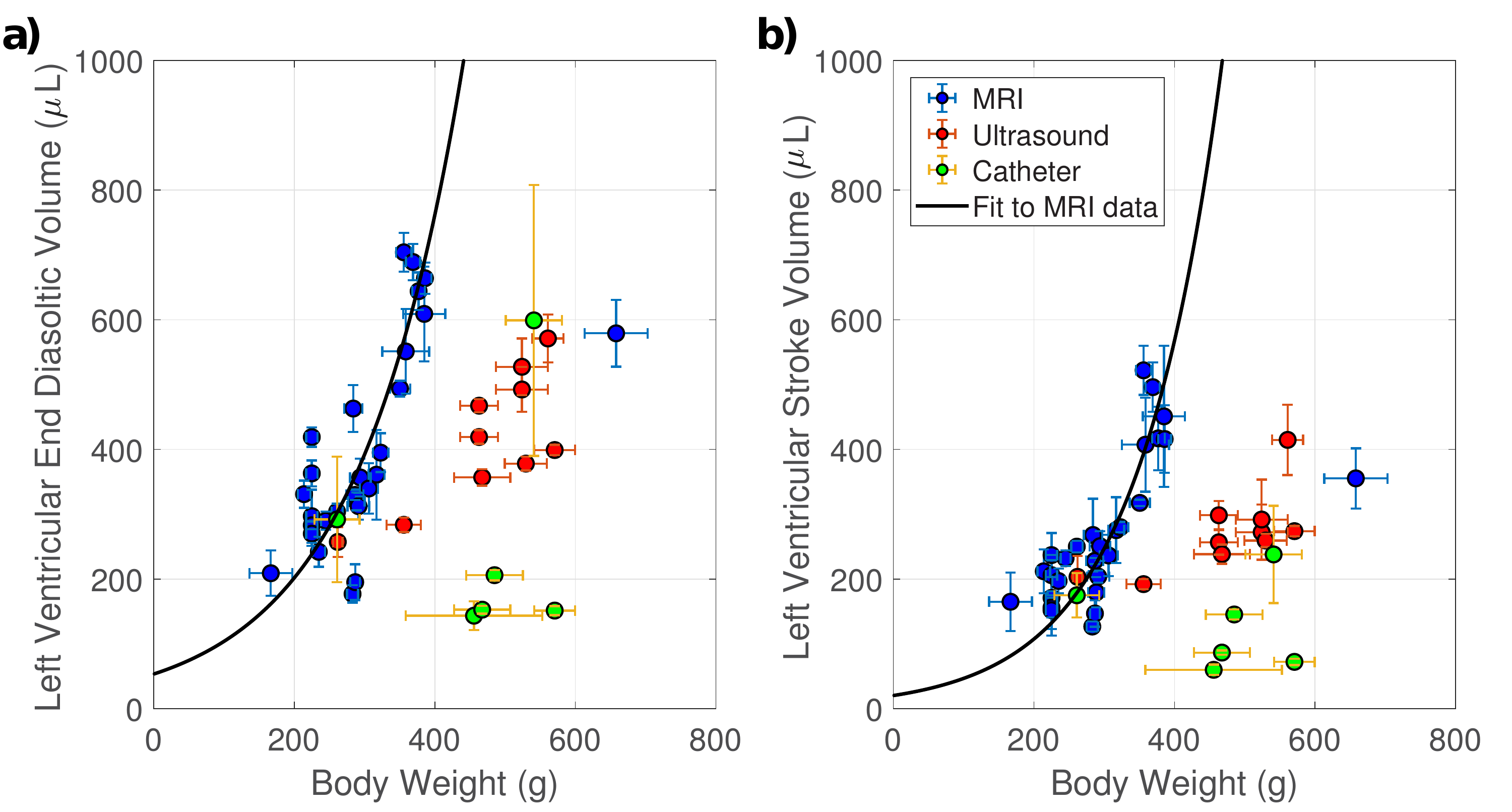}\vspace{-0.4cm}
\caption{Left ventricular EDV ({\bf a}) and SV ({\bf b}) as a function of animal weight and exponential fits (using (\ref{eq:volcal}) to literature data from  catheter~\cite{hwang2009effects,cosyns2007effect,al2002magnetic,holt1968ventricular,nordbeck2013impact,litwin1991induction,engle2009detection,wise1998magnetic,nahrendorf2001serial,vanhove2005reproducibility,bal2005left,nahrendorf2003chronic,radovits2013rat,korkmaz2016left,todica2013positron,carr2008bone,jones2002left,stuckey2008novel,daire2008cine,ruppert2016myocardial}  and MRI-based measurement modalities~\cite{cosyns2007effect,al2002magnetic,nordbeck2013impact,engle2009detection,wise1998magnetic,nahrendorf2001serial,vanhove2005reproducibility,nahrendorf2003chronic,todica2013positron,carr2008bone,jones2002left,stuckey2008novel,daire2008cine}. Error bars denote the standard deviation from each study.}\label{fig:dataproc}
\end{figure}

Conduction catheters measure blood volume indirectly by measuring the conductance through the changing volume of blood in the left ventricle during the experiment~\cite{baan1981continuous}. This method traditionally requires the infusion of a hypertonic saline solution and a cuvette calibration at the end of the experiment to convert the raw voltage time series data into a left ventricular volume time series. However, in these experiments, perturbations performed between the saline calibration and cuvette measures caused this calibration approach to be inaccurate. So to obtain volumes with a standard physiological range, the cuvette measures were scaled against literature estimates pooling end diastolic and stroke volume data from 20 previously published studies using conduction catheters, ultrasound, and magnetic resonance imaging (MRI)  of the left ventricle in SD, Wistar-Kyoto, and Lewis rat strains (from 191 animals total). These data are shown as a function of animal body weight in Figure~\ref{fig:dataproc}. Since MRI based volume measurement techniques have been adopted as the gold-standard for ventricular volume studies~\cite{kjaergaard2006evaluation}, one can observe that measurements from non-MRI methods tend to under estimate both the end diastolic and stroke volume. To obtain realistic volumes, we fit 13 MRI-based end diastolic and stroke volume data sets to body weight using exponential functions
\begin{equation}
\text{EDV} = a\exp\left(b\left(\frac{\text{BW}}{250\text{g}}-1\right)\right),\ \text{and}\ \text{SV} = c\exp\left(d\left(\frac{\text{BW}}{250\text{g}}-1\right)\right),
\label{eq:volcal}
\end{equation}
where BW is the body weight, EDV is the end diastolic volume, SV is the stroke volume, and $\{a,b,c,d\}$ are estimated parameters.  Predicted values of EDV and SV from (\ref{eq:volcal}) are calculated for each animal in our study. The raw conduction (volume) signal is recorded in volts, therefor the predicted EDV is used to convert the peak voltage for each cardiac cycle to the maximum volume and the predicted SV is used to scale the amplitude of the voltage signal.

\subsection{Model}
\label{Subsection:Model}

\begin{figure}[h!]
\centering
\includegraphics[width = .65\textwidth]{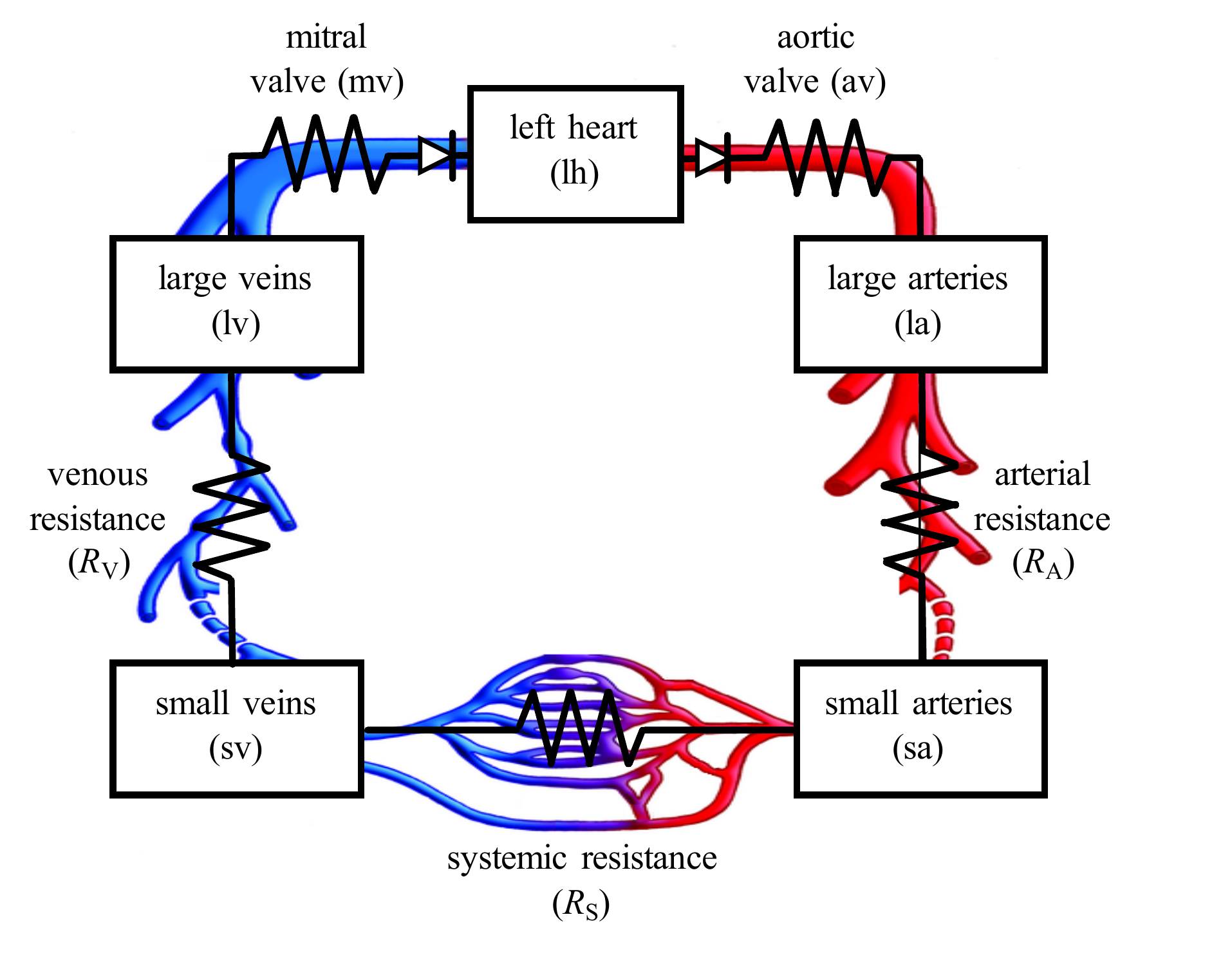}\vspace{-0.4cm}
\caption{Systemic circulation represented using five compartments, including the left heart (lh) (e.g., the left ventricle), the large (la) and small (sa) systemic arteries, and the large (lv) and small (sv) systemic veins. The model is analogous to an RC circuit with capacitors denoting vessel elastance and resistors separating each compartment. Pumping of the heart is ensured by a time-varying elastance function \eqref{eq:elast}. }
\label{fig:circuit}
\end{figure}

Similar to previous studies~\cite{Williams14}, we use a five-compartment model to predict pressure, flow, and volume in  the left ventricle, the large and small arteries and veins; see Figure~\ref{fig:circuit}. Using an electrical circuit analogy, blood flow ($q$) is analogous to current, pressure ($p$) to voltage, volume ($V$) to charge, vessel resistance ($R$) to electric resistance, and vessel elastance ($E$) to the reciprocal of capacitance.  For each compartment, dynamics are predicted from three equations relating pressure, flow and volume. The flow in and out of each compartment is proportional to pressure via Ohm's law 
\begin{equation}
 q = \frac{p_\text{in} - p_\text{out}}{R};
 \label{eq:ohm}
\end{equation}
the pressure and volume in each compartment is related to elastance via
\begin{equation}
 p-p_{\text{ext}} = E(V-V_{\text{un}}) = E V_{str}, 
\label{eq:presVol}
\end{equation}
where $p_\text{ext}$ is the external tissue pressure, $V_\text{un}$ is the unstressed blood volume (both assumed constant), and $V_{str}$ is the stressed blood volume; and conservation of volume gives
\begin{equation}
\frac{dV}{dt} =\frac{dV_{str}}{dt} = q_\text{in} - q_\text{out}.
\label{eq:VolCons}
\end{equation}
For the circuit shown in Figure~\ref{fig:circuit}, we derive a system of five differential equations for the stressed volume of the form (\ref{eq:VolCons}), detailed in the Appendix.  

\begin{figure}[h!]
\centering\includegraphics[scale=0.35]{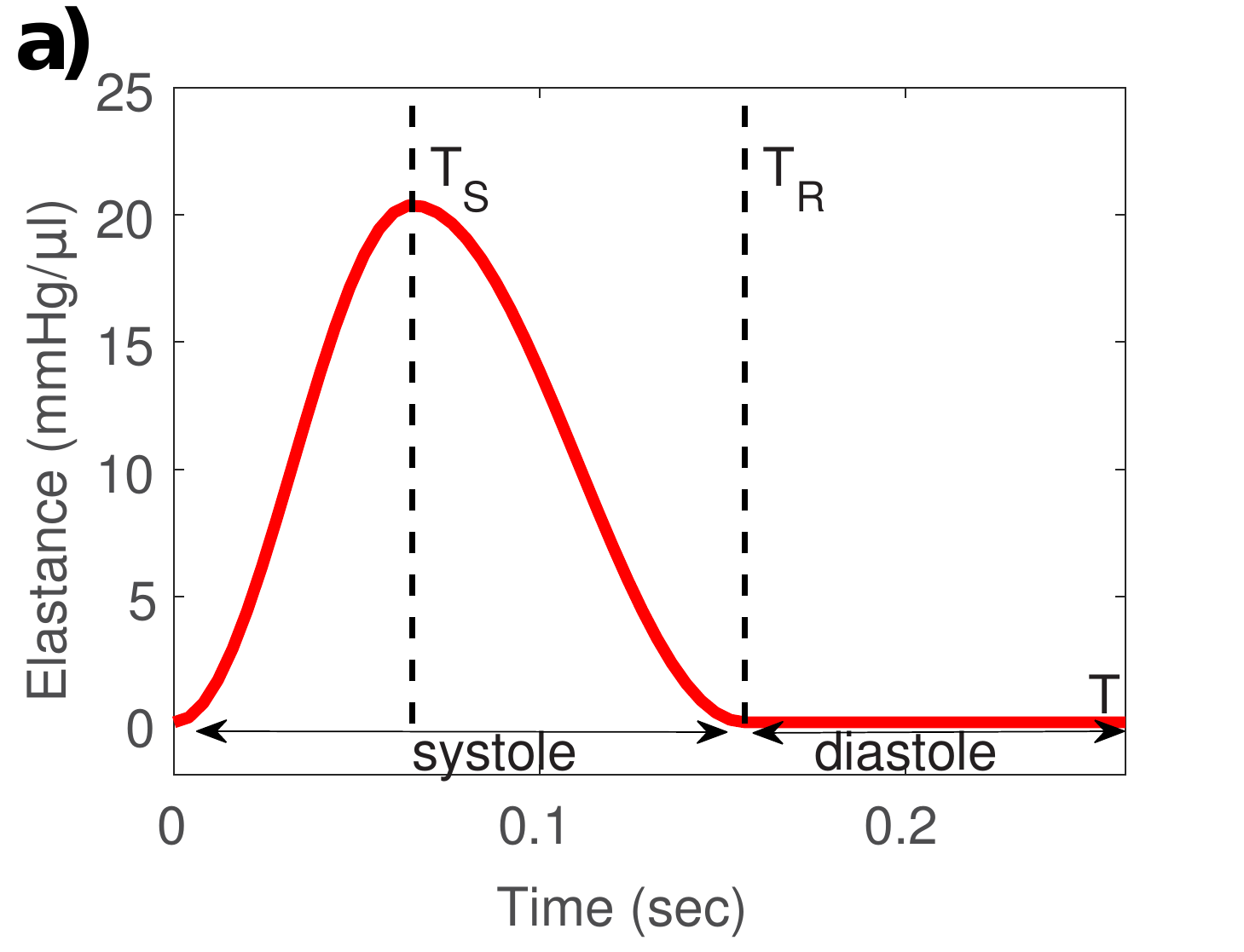}\hspace{-0.3cm}
\includegraphics[scale=0.25]{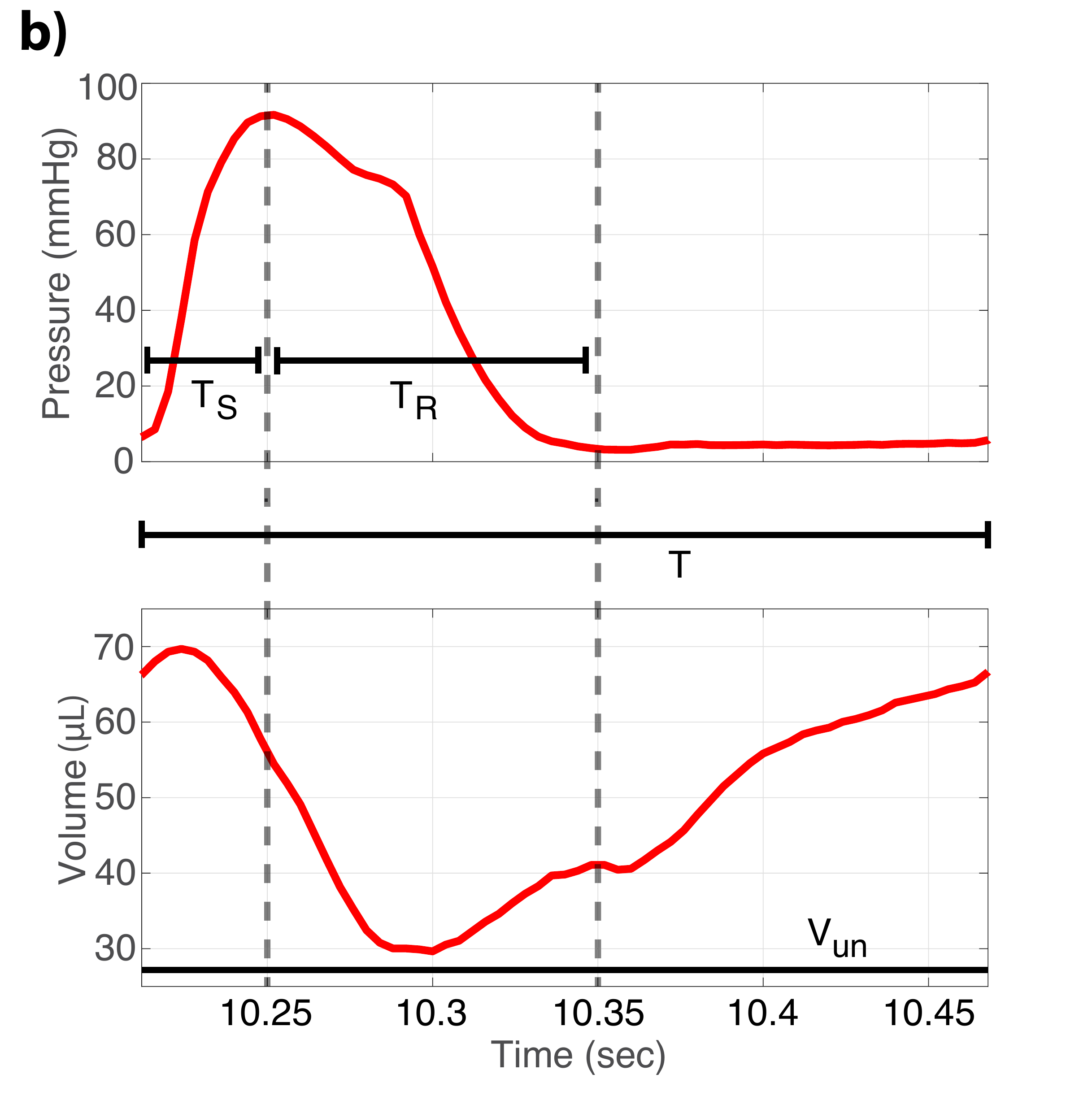}\vspace{-0.4cm}
\caption{The time-varying elastance is modeled using a smooth piecewise trigonometric function defined over the length of one cardiac cycle $T$ (given by equation (\ref{eq:elast})). Maximum elastance $E_\text{M}$ is achieved at $t=T_S$ (at the end of systole) and the end contraction is marked by $t=T_R$. Plots of the elastance function ({\bf a}) and representative pressure and volume data ({\bf b}) are shown to illustrate how the timing parameters $T_{\text{S}}$ and $T_{\text{R}}$ are estimated (see Section~\ref{Subsection:NominalParameters}).}
\label{fig:elastance}
\end{figure}

The beating of the heart (Figure~\ref{fig:elastance}) is modeled by a periodic time-varying elastance function~\cite{Ellwein08}  defined over one cardiac cycle of length $T$ as
\begin{equation}
  E_\text{lv}(t) = \left\{ \begin{array}{ll}
       \displaystyle   E_\text{m} + \frac{E_\text{M}-E_\text{m}}{2} (1-\cos (\pi t/T_\text{S})) &  0 < t < T_\text{S}  \\[1em]
       \displaystyle   E_\text{m} +\frac{E_\text{M}-E_\text{m}}{2}  \cos\left(\pi(t-T_\text{S} )/(T_\text{R}-T_\text{S})\right) &T_\text{S}<t<T_\text{R} \\[1em]
       \displaystyle   E_\text{m}     & T_\text{R} < t < T, 
       \end{array} \right.
       \label{eq:elast}
\end{equation}
where $E_\text{m}$ and $E_\text{M}$ denote the minimum and maximum elastance, respectively, of the left ventricle.  $T_\text{S}$ denotes time for end systole and $T_\text{R}$ the time at which the heart has relaxed to its diastolic value.
We model the two heart valves using its electrical equivalent, a diode, i.e.
\begin{equation}
   q_\text{valve} = \left\{ \begin{array}{ll} 
       \displaystyle \frac{p_\text{in} - p_\text{out}}{R_\text{valve}}  &  \mbox{if } \  p_\text{in} > p_\text{out} \mbox{ (e.g., if valve is open)}   \\[1em]
       \ \ \ \ \ 0                                                   & \mbox{otherwise (e.g., if valve is closed),}
   \end{array} \right.
\end{equation}
where $\text{valve} = \text{av or mv}$, representing the aortic (av) and mitral (mv) valves, respectively.

In summary, the model takes the form
\begin{eqnarray*}
\frac{dx}{dt} &=& f(x,t;\theta),\ \ \ x(0) = x_0\\
       x &=& \left\{V_\text{lh}, V_\text{la}, V_\text{sa}, V_\text{sv}, V_\text{lv} \right\}\\
\theta &=& \left\{R_\text{A}, R_\text{S},  R_\text{V},  E_\text{la},  E_\text{sa}, 
                         E_\text{sv}, E_\text{lh},  T_\text{S},  T_\text{R},   E_\text{m}, 
                         E_\text{M} \right\} \\
      y &=& \left\{V_\text{lh}, P_\text{lh} \right\},
\end{eqnarray*} 
where $x$ denotes the model states, $\theta$ the model parameters, and $y$ the model output.

\subsection{Nominal parameters and initial values} 
\label{Subsection:NominalParameters}

Nominal parameters and initial conditions for the ODEs can be obtained from analysis of data and known physiological features extracted from literature. In this study, parameters include resistance, elastance, and timing of the cardiac cycle, which are determined as a function of resting blood volumes, pressures, cardiac output, and heart rate. In the following sections we discuss how to calculate \textit{a priori} values for all model parameters,  listed in Table~\ref{tab:nompar}.

\subsubsection{Blood volume}

Total blood volume for healthy adult Wistar rats is 57$\mu$l/g body weight~\cite{trippodo} (listed in Table~\ref{tab:data}). Measurements analyzed in this study are from 2 male and 1 female healthy Sprague Dawley rats (Charles River Laboratories, USA).   Following suggestions by Young~\cite{young} and Gelman~\cite{gelman}, the total volume for the purpose of nominal value estimation is distributed with 2.5\% in the large arteries, 7.5\% in the large veins, 20\% in the small arteries  and 70\% in the small veins to 70\%, i.e. for each compartment, the volume is given by
\begin{equation}
   V_i = d_i V_\text{total},
\end{equation}
where $V_\text{total}$ is the total blood volume, $V_i$ refers to the $i$th compartment volume, and $d_i$ denotes the percentage of the total blood volume to compartment $i$.  

As noted above, the model is formulated in stressed volume, which is a small percentage of the total blood volume.  Literature estimates for the total stressed volume vary significantly, from about 10\% to 40\%~\cite{beneken,trippodo,gelman,young,Gotwals00}. We found no resources describing distribution of stressed and unstressed volume across organs or between arteries and veins in rats, therefore we  used the same stressed volume fraction (30\%) in all compartments, i.e.
\begin{equation}
  V_{i,\text{str}} = 0.3\, V_i, \ \ \ \text{for all } i,
\end{equation}
where $i$ denotes "left heart" (lh), "large arteries" (la), "small arteries" (sa), "small veins" (sv), or "large veins" (lv).

\subsubsection{Pressure}

Measurements of left ventricular pressure is used to approximate pressures in all vascular compartments. Starting at the large arteries, the aortic valve open when the pressure in the left ventricle exceeds the aortic pressure. The maximum pressure in the  systemic arteries follow the one in the left ventricle. Assuming a constant pressure drop from left ventricle through the small arteries we let 
\[
  p_{\text{la,M}} = 0.99 \max(p_{\text{lh,d}}), \ \ \  p_{\text{sa,M}} = 0.99  p_{\text{la,M}},
\]
where $p_{\text{lh,d}}$ denotes the left ventricular pressure data, $p_{\text{la,M}}$ denotes maximum larger artery pressure, and $p_{\text{sa,M}}$ denotes maximum small artery pressure. Without pressure measurements in the systemic arteries, more assumptions are needed to approximate the mean arterial pressure. Assuming a pulse pressure of $p_{a,pulse} = 30$ mmHg, large and small artery diastolic pressure can be estimated as $p_{\text{la,dia}}= p_{\text{la,M}}- p_{a,pulse}$, and $p_{\text{sa,dia}}= p_{\text{sa,M}}-p_{a,pulse}$ from which we can estimate mean arterial blood pressure  as
\[
  \overline{p}_{\text{la}} = p_{\text{la,dia}} + 1/3 \, p_{\text{a,pulse}}, \ \ \ \overline{p}_{\text{sa}} =  p_{\text{sa,dia}} + 1/3 \, p_{\text{a,pulse}}
\]
using the common clinical approximation~\cite{KlabundeWeb}.

This model does not represent the right ventricle or pulmonary vasculature so the pressure in the systemic veins becomes the filling pressure for the left ventricle. Thus similar to the arterial side, we assume a 10\% pressure drop between large and small veins, so we let
\[
  \overline{p}_{\text{lv}} = 1.1 \min (p_{\text{lh,d}}), \ \ \  \overline{p}_{\text{sv}} = 1.1 \overline{p}_{\text{lv}}, 
\]

\subsubsection{Cardiac output}

For each rat, the average stroke volume (Table~\ref{tab:data}) is extracted from measurements of maximum and minimum left ventricular volume. Cardiac output can subsequently be calculated as the product of stroke volume and heart rate (i.e. $\text{CO} = \text{HR} \ V_\text{stroke} $).

\subsubsection{Vascular resistance}

Using the baseline cardiac output and pressure, the vascular resistances can be computed from Ohm's law (\ref{eq:ohm}). For example, arterial resistance is predicted as
\begin{equation}
  R_\text{A} = \frac{\overline{p}_{\text{la}} - \overline{p}_{\text{sa}}}{\text{CO}}. \label{eq:Rs}
\end{equation}

Instead of using the mean pressures to calculate valve resistance we consider the maximum and minimum pressure to estimate the resistance across the aortic and mitral which can then are expressed as
\begin{equation}
    R_\text{av} = \frac{\max(p_{\text{lh,d}})  - p_\text{la,M}}{\text{CO}} \qquad \mbox{ and } \qquad
    R_\text{mv} = \frac{\overline{p}_{\text{lv}}  - \min(p_\text{lh,d})}{\text{CO}}, 
    \label{eq:Rv}
\end{equation}
where $R_{\text{av}}$ is the aortic valve resistance and $R_{\text{mv}}$ is the mitral valve resistance.  Since the venous pressure do not vary significantly, the average pressure of the large veins is used to predict $R_{\text{mv}}$.

\subsubsection{Vascular compliance}

Each vascular compartment is associated with an elastance (constant). Assuming that the tissue pressure is negligible (e.g. $p_{\text{ext}} = 0$), elastance can be estimated as
\begin{equation}
   E_\text{la} = \frac{p_\text{la,M}}{V_\text{la} -V_{\text{la,un}}}, \label{eq:C}
\end{equation}
following the pressure-volume relation (\ref{eq:presVol}), set up for the large arteries as a representative example.

\subsubsection{Heart parameters}

The elastance function (\ref{eq:elast}) has four parameters, including timing parameters denoting the length of the cardiac contraction ($T_\text{S}$) and relaxation ($T_\text{R}$), as well as the minimum ($E_\text{m}$) and maximum ($E_\text{M}$) elastance.  Rat unstressed ventricular volume has been estimated at 37$\mu$l~\cite{london2012}.  For each rat, the timing parameters $T_{\text S}$ and $T_\text{R}$ can be extracted from data, as shown in Figure~\ref{fig:elastance}. $T_\text{S}$ denotes the time at which the left ventricular volume reaches it maximum (at $\max(p_{\text{lh,d}})$), and $T_\text{R}$ is the time at which $p_\text{lh}$ reaches its baseline after contraction.  The minimum elastance $E_\text{m}$ is associated with end-diastole, where the left ventricular pressure is minimal and ventricular volume is maximal, and the maximum elastance $E_\text{M}$ is associate with systole, where the ventricular pressure is  maximal and ventricular volume is minimal. These considerations let us set
\begin{equation}
     E_\text{m} = \frac{\min(p_{\text{lh,d}})}{\max(V_{\text{lh,d}}) - V_\text{lh,un}}  \qquad \mbox{ and } \qquad  
     E_\text{M} = \frac{\max(p_{\text{lh,d}})}{\min(V_{\text{lh,d}}) - V_\text{lh,un}}.  \label{eq:EMm}
\end{equation}

\subsubsection{Initial conditions}                

Assuming that the model simulation begins at the end of systolic contraction (beginning of diastolic filling), we set the initial value of the volume in each compartment to their stressed volume ($V_{i,\text{s}}$). This implies that  
\begin{equation}
V_{0,i} = V_{i,\text{s}} = 0.3V_{i}
\end{equation}
where $V_{0,i}$ is the initial volume of the $i$th compartment.

\begin{table}[t!]
\caption{Quantities used to determine nominal parameter estimates.} \label{tab:nompar}
{\renewcommand{\arraystretch}{1.4}
{\footnotesize 
\begin{tabular}{l l l |l l l} \hline 
Quantity & Equation & Reference &Quantity & Equation & Reference\\ \hline
$V_\text{la}$   & $0.025 \, V_\text{total}$ & \cite{young,gelman}     &  $p_\text{la,M}$             & $0.99 \, \max(p_\text{lh,d})$  & data\\
 & & & $\overline{p}_\text{la}$ & $p_\text{la,dia} + 1/3 \, p_\text{la,pulse}$ & \cite{KlabundeWeb}\\
$V_\text{sa}$  & $0.2 \,V_\text{total}$      & \cite{young,gelman}     &  $p_\text{sa,M}$            & $0.99 \,p_\text{la,M}$            & data \\
  & & &$\overline{p}_\text{sa}$ & $p_\text{sa,dia} + 1/3 \, p_\text{sa,pulse}$ & \cite{KlabundeWeb}\\ 
$V_\text{lv}$   & $0.075 \,V_\text{total}$  & \cite{young,gelman}     & $\overline{p}_\text{lv}$   & $1.1 \, \min(p_\text{lh,d})$    & data\\ 
$V_\text{sv}$  & $0.7 \,V_\text{total}$      & \cite{young,gelman}     & $\overline{p}_\text{sv}$  & $1.1 \, \overline{p}_\text{lv}$ & data\\ \hline
$R_\text{av}$  & $\ds\frac{\max(p_{\text{lh,d}})-p_\text{la,M}}{\text{CO}}$                 & (\ref{eq:Rv}) & $E_\text{la}$   & $\ds\frac{p_{\text{la,M}}}{V_\text{la} - V_{\text{la,un}}}$  & (\ref{eq:C})  \\
$R_\text{mv}$ & $\ds\frac{\overline{p}_\text{lv} - \min(p_{\text{lh,d}})}{\text{CO}}$    & (\ref{eq:Rv}) & $E_\text{sa}$  & $\ds\frac{p_\text{sa,M}}{V_\text{sa}-  V_{\text{sa,un}}}$ &  (\ref{eq:C}) \\
$R_\text{A}$    & $\ds\frac{\overline{p}_\text{la}-\overline{p}_\text{sa}}{\text{CO}}$  & (\ref{eq:Rs}) & $E_\text{lv}$   & $\ds\frac{\bar{p}_{\text{lv}}}{V_\text{sv}-V_{\text{sv,un}}}$   & (\ref{eq:C}) \\
$R_\text{S}$    & $\ds\frac{\overline{p}_\text{sa}-\overline{p}_\text{sv}}{\text{CO}}$   & (\ref{eq:Rs})& $E_\text{sv}$   & $\ds\frac{\bar{p}_\text{sv}}{V_\text{sv}-V_{\text{sv,un}}}$  & (\ref{eq:C}) \\
$R_\text{V}$    & $\ds\frac{\overline{p}_\text{sv}-\overline{p}_\text{lv}}{\text{CO}}$    & (\ref{eq:Rs}) &  &  \\ \hline
$T_{\text S}$   & $t_{\max(V_{lh,d})}$                                                                         &                     &  $E_{\text M}$ & $\frac{\overline{p}_{\text{lv}}}{ \max(V_{\text{lh,d}})-V_{\text{lh,un}}}$& (\ref{eq:EMm}) \\
$T_{\text R}$   & $t_{\min(p_{lh,d})}$                                                                          &                     &$E_{\text m}$ & $\frac{p_{\text{la,dia}}}{ \min(V_{\text{lh,d}})-V_{\text{lh,un}}}$ & (\ref{eq:EMm}) \\[.5em] \hline
\end{tabular}}}
\end{table}

\section{Model Analysis}\label{Section:ParamEst}

The model described in Section~\ref{Subsection:Model} is linear with respect to the states but nonlinear with respect to the parameters. This gives rise to multiple parameter interactions that inevitably complicate the parameter estimation process. While a model with many nonlinear parameter interactions can be structurally identifiable, unidentifiable parameter relationships  often result in an ill-conditioned optimization problem~\cite{ipsen2011rank}.
 
In this section, we outline a systematic methodology for parameter estimation, comprising identification of
\begin{enumerate}
\item{\it Nominal parameters} from literature and available data followed by  a baseline simulation to ensure an appropriate model response. 
For the model analyzed here this step requires two parts: i) using data to compute subject specific nominal parameter values, ii) determine a shift in data to ensure that model predictions and data are in phase.
\item {\it Local  sensitivities} used to study how the  parameters influence the model output.  
\item {\it Structured correlations} used to determine a subset of parameters with minimal parameter interactions.  
\item {\it Parameter estimates} obtained using the Levenberg-Marquardt optimization method, estimating the subset of identifiable parameters minimizing the least squares error between the model output and  available data. 
\item {\it Frequentist prediction and confidence intervals} used to quantify uncertainty in the model solutions.
\item {\it Parameter distributions and credible intervals} obtained using the DRAM (Delayed Rejection Adaptive Metropolis) algorithm.
\end{enumerate}
While the analysis is devised for a relatively simple cardiovascular model with a specific set of output data (left ventricular pressure and volume), the approach introduced here is applicable to any predictive model fitted to time-varying data. Obtaining the nominal parameter of step one is model specific however, steps 2-6 are more generic and the approach presented here can be more generally applied

\subsection{Local Sensitivity Analysis} 

Sensitivity analysis quantifies how the model output changes in response to changes in parameter values~\cite{Ellwein08}. In this study we use derivative-based  sensitivities to quantify the local influence of the model output on each parameter.  Similar to the study by Pope et al.~\cite{Pope09}, we computed  sensitivities from partial derivatives of the model output residual with respect to each model parameter, i.e. we define the sensitivity matrix $\mathsf{S}$ as 
\begin{equation} \label{eq:SensMatrix}
  \mathsf{S}_{i,j} = \frac{\partial r(t_i,\theta)}{\partial \theta_j},  
\end{equation}
where  $\theta_j$ is the $j$th parameter and $t_i$ is the $i$th time step. When estimating parameters in a log-transformed space (\ref{eq:SensMatrix}) can be expressed as
\begin{equation*}
\frac{\partial r(t_i,\theta)}{\partial \ln(\theta_j)} = \theta_j\frac{\partial r(t_i,\theta)}{\partial \theta_j}.
\end{equation*}

The matrix $\mathsf{S}$ can be calculated analytically for simple models; however, numerical approximation of $\mathsf{S}$ is more practical for complex models.  Here we employ finite differences to approximate $\mathsf{S}$ by 
\[
   \mathsf{S}_{i,j} = \frac{ r(t_i,\theta_j+h e_i) - r(t,\theta_j)}{h},
\]
where $h$ is chosen to reflect the precision of the model output and $e_i$ is the unit vector in the $i$th component direction.  If the error in the model evaluation (ODE solver error tolerance) is on the order of $\varepsilon$, the step should be $h=\sqrt{\varepsilon}$ to get an error of the same magnitude in the sensitivities~\cite{Pope09}.  We tested the stability of our finite difference approximation by reducing the ODE\ solver tolerance and observing that the results of the sensitivity analysis converged the same values (results not shown). To get a rough approximation of how a parameter influences model behavior we use ranked sensitivities $R_j$, defined as the two-norm over each column of $\mathsf{S}$
\[
  R_j = \left(\sum^N_{i =1}S_{ij}^2\right)^{1/2}.
\]

\subsection{Subset Selection: Structured Correlation Analysis}
\label{subsection:subset}

The  model considered here has inherent parameter interactions that necessitate the need for selecting parameter subsets with minimal unidentifiable parameter interactions. While various subset selection algorithms exist (e.g.~\cite{Miao11}), we employ the structured correlation method by Ottesen et al.~\cite{OlufsenCorr} to construct a set of identifiable model parameters. According to this method, a pair of parameters with a large correlation (and strongly coupled uncertainty) cannot both be uniquely estimated.  This method systematically removes the least sensitive correlated parameters parameters until an identifiable set remains. The input to this method is the sensitivity matrix - which means that any correlation  is only valid within some neighborhood of the parameter values used to construct the sensitivity matrix.

Using the sensitivity matrix $\mathsf{S}$ in \eqref{eq:SensMatrix}, the covariance matrix $\mathsf{\Gamma}$ is given by the inverse of the Fisher Information Matrix (FIM)
\begin{equation}
 \mathsf{\Gamma} = \mathsf{F}^{-1},
\end{equation}
where $F=\sigma^2\mathsf{S}^T\mathsf{S}$. Note that $\mathsf{F}$ can only be inverted if $\mathsf{S}$ has full rank. Linearly dependent columns of $\mathsf{S}$ are a result of parameters being perfectly correlated, meaning that a parameter can be algebraically expressed in terms of other parameters. Additionally, columns of insensitive parameters ($R_j<h$) can lead to $F$  having a large condition number  and thus should be removed \textit{a priori} from the sensitivity matrix input. The entries of $\mathsf{\Gamma}$ are used to compute the correlation matrix $\mathsf{C}$ with entries given by
\[
\mathsf{C}_{i,j} = \frac{\mathsf{\Gamma}_{i,j}}{\sqrt{\mathsf{\Gamma}_{i,i} \mathsf{\Gamma}_{j,j}}}.
\]
$\mathsf{C}$ is a symmetric matrix with ones along the diagonal. Parameter pairs with a correlation value greater than $\gamma$ (user defined threshold) are denoted correlated. Correlations between parameters show how the parameter values depend on each other when fitting experimental data from the same system.

\subsection{Nonlinear Least Squares Optimization: Levenberg-Marquardt}
\label{section:freqopt}

The goal of nonlinear least squares optimization is to estimate the set of parameters $\hat{\theta}$ that minimizes the difference between the model output $x(t,\theta)$ and the data $y$, assumed to be some function of the model output corrupted with additive noise; e.g. 
\begin{equation}
 y_i = g(x(t_i,\theta)) + \epsilon_i, 
 \label{eq:statmod}
\end{equation}
where $y_i$ denotes the $i$th data point, $x(t_i,\theta)$ the model response at the $i$th time point, $g(\cdot)$ the observation function mapping the model variables to the measured states, and $\epsilon_{i}$ the normally distributed observation error. In this study, $g(\cdot)$ extracts the left ventricular volume and pressure time-series.

To fit our model to the data, we use generalized nonlinear least squares to determine a parameter set $\hat{\theta}$, which minimizes the  sum of squares cost function 
 \begin{equation}
 J(\theta) = \sum_{k=1}^2 J_k(\theta),
\
 J_k(\theta) =r_k^Tr_k.
 \label{eq:cost}
\end{equation}
The residual vectors $r_k$ are defined as
\begin{equation}
r_1 = \frac{p_\text{lv}^\text{m} - p_\text{lv}^\text{d}}{\max(p^\text{d}_{\text{lv}})}  \qquad \mbox{ and } \qquad
r_2 = \frac{V_\text{lv}^\text{m} - V_\text{lv}^\text{d}}{\max(V^\text{d}_{\text{lv}})-\min(V^\text{d}_{\text{lv}})} \ .  \label{eq:res}
\end{equation}
The superscripts ``m'' and ``d'' denote the model and data, respectively. Dividing the residuals by the amplitude of the data ensures that the optimization procedure gives equal weight to both model outputs.    

To estimate $\hat{\theta}$ we employ the Levenberg-Marquardt optimization routine by Kelley~\cite{kelley}. To obtain a well scaled problem, we estimate the natural log of the true parameter values. As part of the optimization routine we discard parameter values that are 20 times larger or smaller than the nominal estimate. Like other gradient based optimization routines, the choice of the initial parameter vector $\theta_0$ is important. If the cost function has multiple minima the optimization routine may end in an undesirable or unrealistic local minimum.  Using the physiologically-justified nominal parameter values described in Section~\ref{Subsection:NominalParameters} helps to place our $\theta_0$ closer to a physiologically realistic minimum. To verify that the optimization converged we randomly perturbed the nominal values by 10\% and observed that the parameters all converged to the same values.

 \subsection{Uncertainty Quantification}

Uncertainty quantification (UQ) is the process of determining uncertainties in estimated model parameters given uncertainties in the model formulation and experimental measurements (the inverse problem), as well as establishing how uncertainties in model inputs (such as parameters) affect the model output (forward propagation of uncertainty).  In this study, we utilize UQ procedures from both frequentist and Bayesian statistical frameworks.  We calculate confidence intervals and Bayesian credible intervals to measure the precision of the model in predicting the mean response.  These approaches are outlined below; for more details, see~\cite{SmithUQ}.

\subsubsection{Frequentist Approach}
\label{Subsubsection:Freq}

Frequentist uncertainty propagation methods are computationally inexpensive compared to their Bayesian analog. Most frequentist statistics are derived from asymptotic assumptions assuming that the uncertainty distributions take a Gaussian shape, whereas Bayesian methods make no assumption about the shape of the underlying distributions. One of the main benefits of the asymptotic assumptions is that uncertainty distributions can be expressed as explicit formulas in the frequentist perspective. Frequentist confidence intervals can be calculated from
\begin{equation}
 \hat{y}_i \pm t_{N-p}^{\alpha/2} s(g_i^T (\mathsf{S}^T \mathsf{S})^{-1} g_i)^{1/2},
 \label{eq:conf}
\end{equation}
where  $t_{N-p}^{\alpha/2}$ is the student t-distribution with  $N-p$ degrees of freedom ($N$ is the number of data points and $p$ is the number of parameters),  $s$ is the estimate of the model standard deviation $\sigma$, $\mathsf{S}$ is the sensitivity matrix, and $g_i$ is the $i$th row of $\mathsf{S}$ stacked as a column vector.  Frequentist prediction intervals can be calculated in a similar manner by 
\begin{equation}
 \hat{y}_i \pm t_{N-p}^{\alpha/2} s(1 + g_i^T (\mathsf{S}^T \mathsf{S})^{-1} g_i)^{1/2}.
\label{eq:freq}
\end{equation}

\subsubsection{Bayesian Approach: Delayed Rejection Adaptive Metropolis (DRAM)}
\label{Subsubsection:DRAM}

While frequentist methodology is fundamentally rooted in quantifying uncertainty in terms of repeating the data generating procedure, Bayesian inference is conditioned on a single data set; this allows for uncertainty about parameters to be expressed by probability distributions. In the Bayesian framework, $\theta$ represents a vector of random variables. Given observations $y = \{y_1, \dots, y_n\}$, Bayes' formula
\begin{equation}
  \pi(\theta \mid y) = \frac{\pi(y \mid \theta)\pi(\theta)}{\pi(y)}
\end{equation}
describes how the posterior density $\pi(\theta \mid y)$ relates to the prior density $\pi(\theta)$, encompassing any a priori information known about the parameters, and the likelihood $\pi(y \mid \theta)$ of observing the data $y$ for the model given $\theta$.  The marginal density $\pi(y)$ of the data typically functions as a normalization factor and can be determined by 
\begin{equation}\label{eq:normalization}
  \pi(y) = \int \pi(y\mid\pi) \pi(\theta)\, d\theta.
\end{equation}
Under the hypothesis (\ref{eq:statmod}), the likelihood function is given by
\begin{equation}
  \pi_k(y\mid\theta) = \frac{e^{-J_{k}(\theta)/2\sigma_k^2}}{(2\pi\sigma^2)^{n/2}},
\label{eq:likelihood}
\end{equation}
where $J_{k}(\theta)$ denotes the least square cost defined by (\ref{eq:cost}), $n$ is the number of data points, and $\sigma_k^2$ is the model variance - the $k$ index denotes the variance for pressure ($k=1$) or volume ($k =2$). Both $\sigma^2_1$ and $\sigma^2_2$ are parameters represented by probability distributions that are also estimated. With a known likelihood and prior density $\pi_{k}(\theta)$, it is possible to estimate the posterior density $\pi(\theta \mid y)$ if the integral \eqref{eq:normalization} in the normalizing constant can be estimated.  While this route is possible, the evaluation of high-dimensional integrals is a difficult and expensive, and is currently an active area of research; see, e.g., sparse grid methods~\cite{smolyak1963quadrature, bungartz2004sparse} and quasi-Monte Carlo methods~\cite{halton1960efficiency, joe2003remark}.

An alternative is to use Monte Carlo integration to randomly sample from the density $\pi(y\mid\theta)\pi(\theta)$. Many suitable Markov chain Monte Carlo (MCMC) methods exist in the literature (see~\cite{Andrieu2008} for an overview). This study uses the DRAM algorithm~\cite{haarioDRAM}, which combines two methods for improving efficiency of Metropolis-Hastings type MCMC algorithms: delayed rejection (DR)~\cite{MiraDR2001} and adaptive Metropolis (AM)~\cite{HaarioAM2001}.  These Metropolis-type methods are acceptance-rejection algorithms that accept new parameter samples only if the likelihood of the new candidate is higher than the current sample.  DR allows for additional proposals per step if the initially proposed step is not accepted, thereby increasing the acceptance rate and well-mixing of the sample.  AM allows for updating of the covariance matrix based on the history of the sample, thereby helping the algorithm to make better proposals and move toward the correct posterior distribution faster, reducing the burn-in period.

 We use (\ref{eq:likelihood}) as our likelihood distribution, uninformed diffuse Gaussian distributions centered around the optimized point estimate from Section~\ref{section:freqopt} as our prior distribution, and a Gaussian proposal distribution. To ensure a well scaled problem we have DRAM estimate the natural log of the true parameter densities like we do Section~\ref{section:freqopt}. Samples are taken from the DRAM-estimated parameter probability distributions to compute Bayesian credible and prediction intervals; for more details, see~\cite{haarioDRAM, SmithUQ}.  In this study, we utilize the MCMC toolbox by Haario et al. (2006) at {\tt http://helios.fmi.fi/$\sim$lainema/mcmc/}, which includes code for running DRAM as well as for computing Bayesian credible and prediction intervals.

\section{Results}\label{section:results}

In this section, we use the step-by-step procedure described in Section~\ref{Section:ParamEst} to systematically estimate an identifiable subset of parameters for the cardiovascular compartment model derived in Section~\ref{Subsection:Model}  that fit left ventricular volume and blood pressure data described in Section~\ref{Subsection:ExpData}.

\subsection{Nominal parameter values}

\begin{figure}[h!]
\centering
\subfigure[]{ \hspace{-0.5cm} \includegraphics[height=3.7cm]{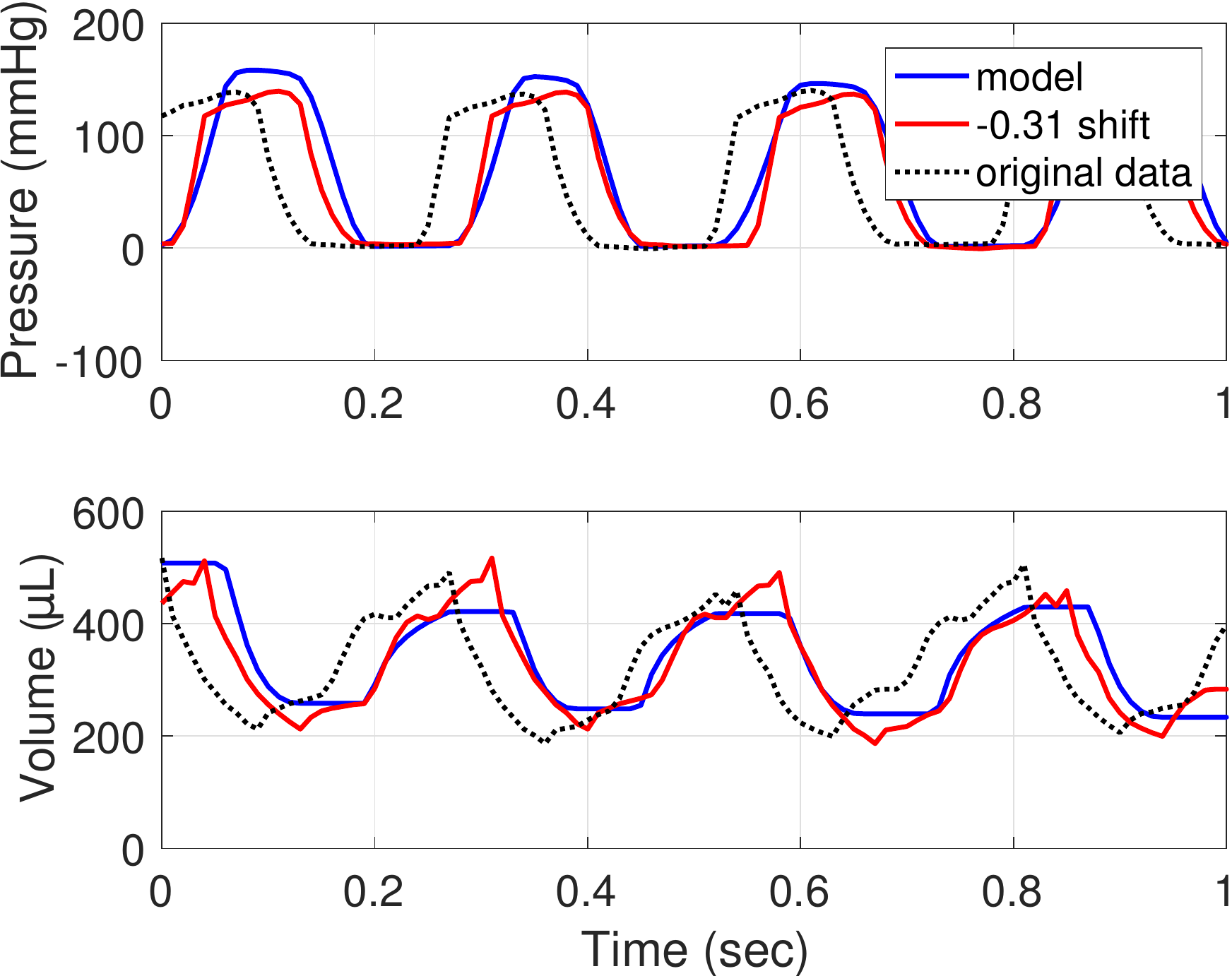}\hspace{-0.13cm}}
\subfigure[]{\includegraphics[height=3.7cm]{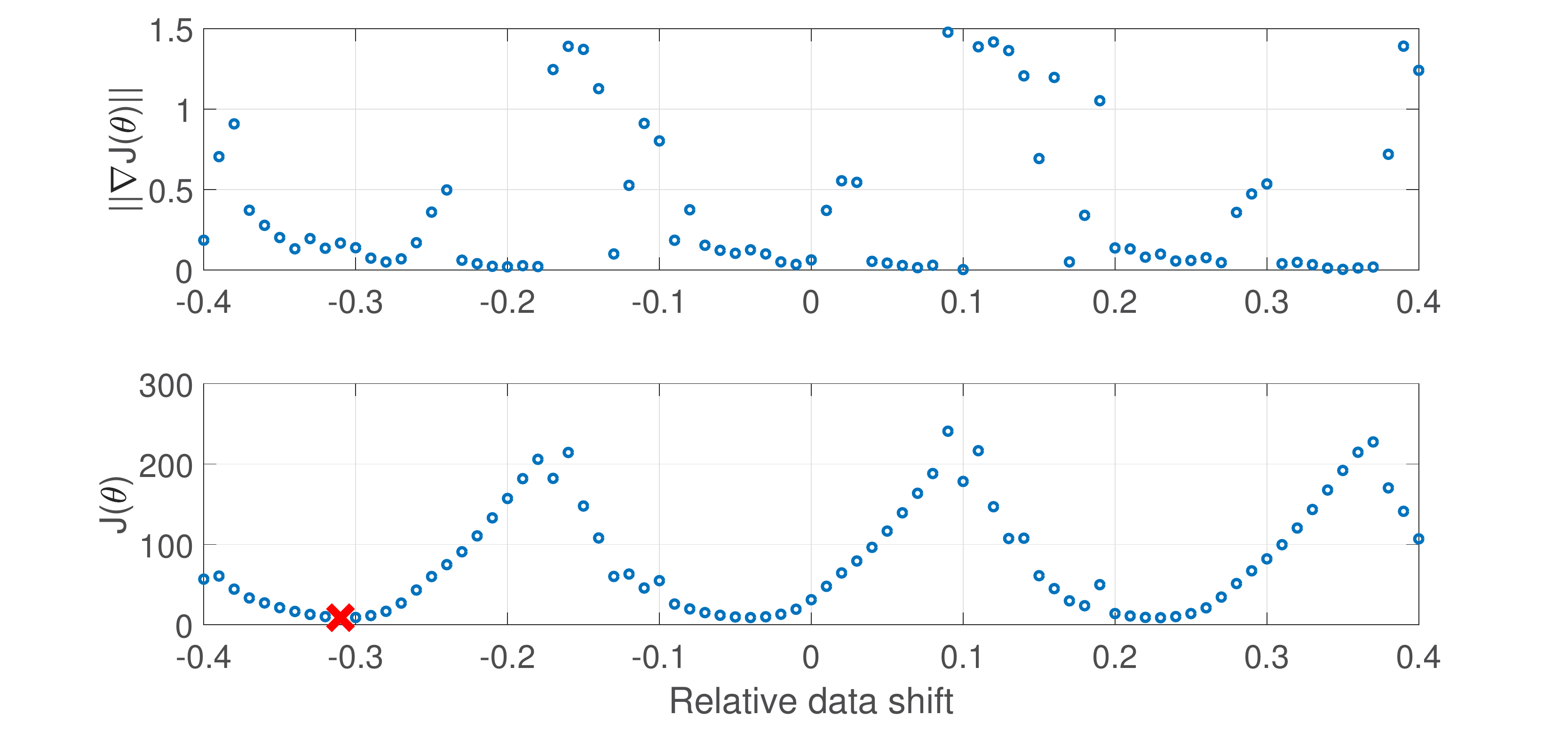}\hspace{-0.5cm} }
\caption{(a) Manual shifting of the data starting point for Rat 1. The blue curve is the model solved using nominal parameters, the dotted curve is the original phase of the data, and the red curve is the same data shifted 0.24 seconds to the right in order align the phase of the model and data. Manual shifting for Rat 2 and 3 data sets are similar. (b) Determining the optimal data shift for Rat 1. The top plot shows the value of the gradient and the bottom plot shows the value of the cost function (\ref{eq:cost}) as a function of relative data shifts. The red ``x'' in the bottom plot denotes the smallest value of the cost function and our optimal data shift. Results for Rat 2 and 3 show a similar pattern.}\label{fig:shiftI}
\end{figure}

Given that sensitivity and subset selection analysis are determined at an estimate of the local minimum, the first step involves using available data and literature to determine the best possible nominal parameter values, as described in Section~\ref{Subsection:NominalParameters}. Second, we manually shift the starting point of the data  so the model predictions and data  both begin at the same point of the cardiac cycle, as shown in Figure~\ref{fig:shiftI}(a) and described in Section~\ref{Section:ParamEst}. This estimate of the data shift is refined in a later model optimization step. 

\subsection{Sensitivity Analysis and Subset Selection}

Next, we employ local sensitivity analysis and structured correlation, along with physiological knowledge, to construct a subset of identifiable model parameters.

\begin{figure}[t!]
\centering\includegraphics[width = 0.8\textwidth]{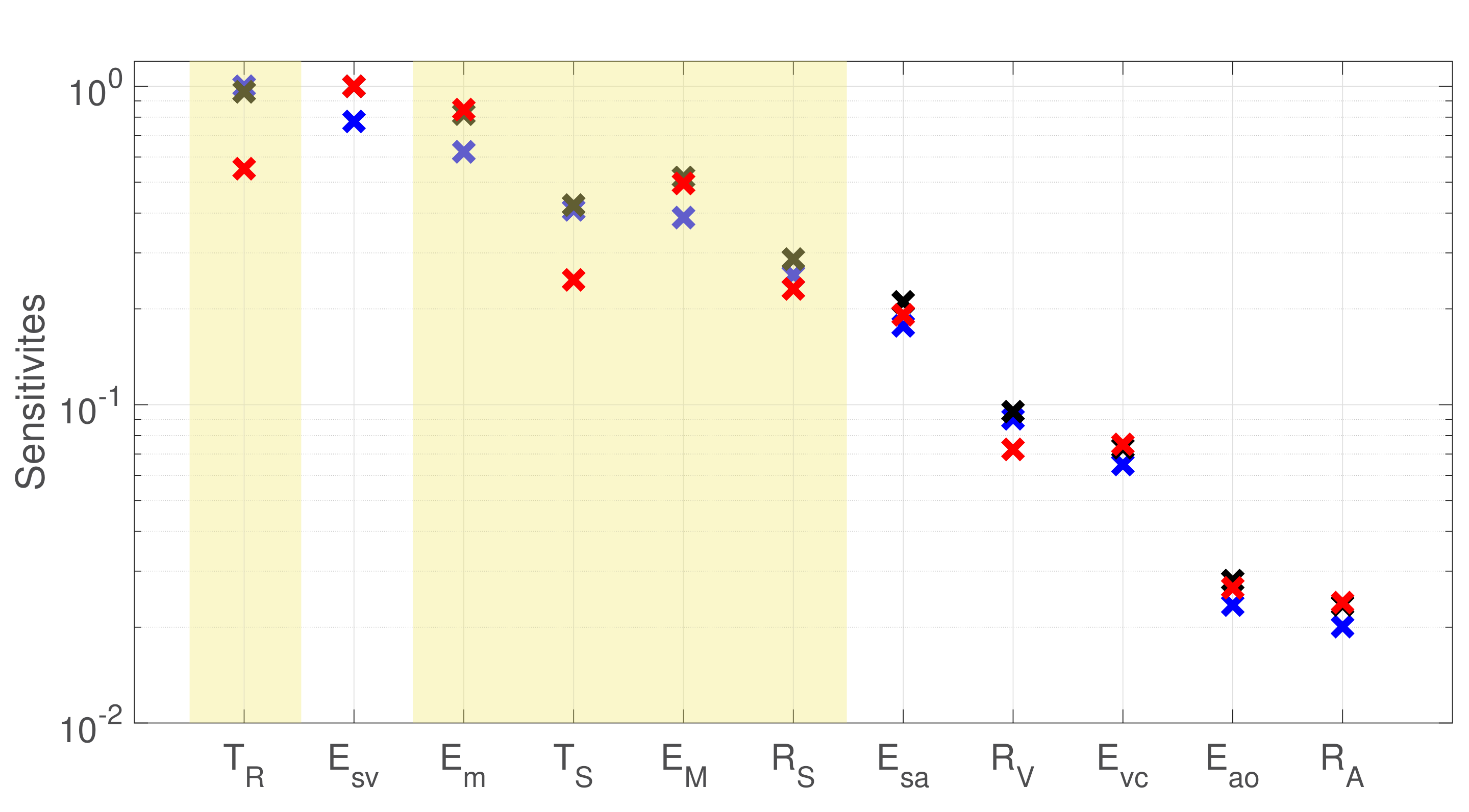}\vspace{-0.4cm}
\caption{Ranked parameter sensitivities. Blue, black, and red x's indicate relative sensitivities of Rats 1, 2, and 3 respectively. Yellow highlighted regions indicate parameters included in the subset (\ref{eq:subset}).}
\label{fig:rankedsens}
\end{figure}

Figure~\ref{fig:rankedsens} depicts normalized ranked parameter sensitivities and the subset of identifiable parameters (\ref{eq:subset}) for all three animals. Ranking the parameters by sensitivity provides  initial insight into what parameters are most influential in determining model behavior. Note that the ranked sensitivities of parameters $R_\text{a}$ and $E_\text{ao}$ are orders of magnitude smaller than the other parameters, thus these parameters were removed \textit{a priori} from the set of parameters analyzed. We used the structured correlation algorithm from Section \ref{subsection:subset} to obtain an identifiable parameter set.  For all three rats we used a correlation threshold $\gamma=0.85$ as an upper bound on the pairwise correlations between parameters and found the identifiable subset
\begin{equation}
  \theta = \{R_\text{s},T_{\text{S}}, T_\text{R}, E_\text{m}, E_\text{M}\}.
\label{eq:subset}
\end{equation}

\subsection{Optimization}
\label{section:opt}

Using the nominal parameters listed in Table~\ref{tab:optpar}, we optimize the parameter subset (\ref{eq:subset}) using a Levenberg-Marquardt optimization scheme; we tested convergence of the optimizer by varying the initial guess around the nominal values.  Results were verified by randomly perturbing the nominal values by 10\% and optimizing these perturbed parameters.  For ten unique perturbations, all of the parameters converged to the same values (results not included).  To ensure that the solution has settled to steady state, computations were done over entire 20 seconds of data, with the least squares cost evaluated over the last 0.5 seconds of the data shown in Figure~\ref{fig:data}. 
As part of the optimization procedure we also optimize the data shift which was manually determined in step 1. This was done by repeating optimizations for a large number of shifts. Results showed  (\ref{fig:shiftI}(b)) that there are multiple parabolic minimums for the cost reflecting the pulsatile nature of cardiovascular mechanics. We chose the minima that provided the smallest cost (\ref{eq:cost}). Results depicted in Figure~\ref{fig:shiftI} shows the cost (\ref{eq:cost}) and gradient plotted as a function of relative data shifts for Rat 1 (similar results were obtained for rats 2 and 3). 

Optimization results for all three rats, with optimal shifts, are reported in Table~\ref{tab:optpar} listing the nominal and optimized parameter values, and  Figure~\ref{fig:UQ}, showing the model fit with optimized parameter values compared to the data. Note that we report the true parameter values and not the natural log of the parameters used by the Levenberg-Marquardt optimizer.

\subsection{Uncertainty Quantification}

The point-estimates obtained using the Levenberg-Marquart routine were used to initialize both frequentist and Bayesian UQ methods,  to construct confidence and credibility intervals along with prediction intervals for the model output predictions. This choice was motivated by wanting to minimize the already high computational cost of solving stiff ODEs. To demonstrate our methodology is reproducible across data sets we repeated computations for all three data sets shown in  Figure~\ref{fig:data}. 

\begin{figure}[h!]
\centerline{
\includegraphics[height = 4cm]{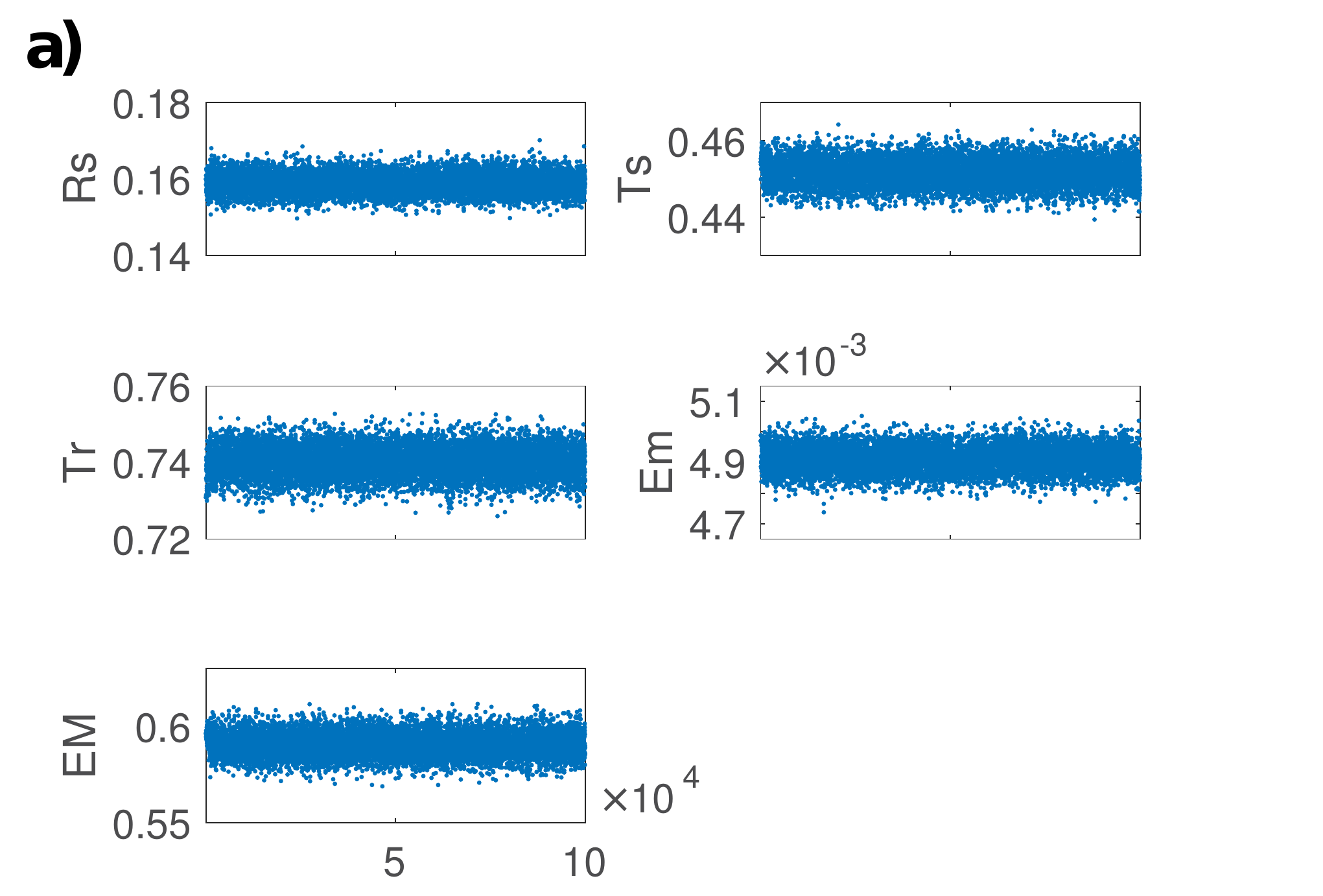}\hspace{-.75cm}
\includegraphics[height = 4cm]{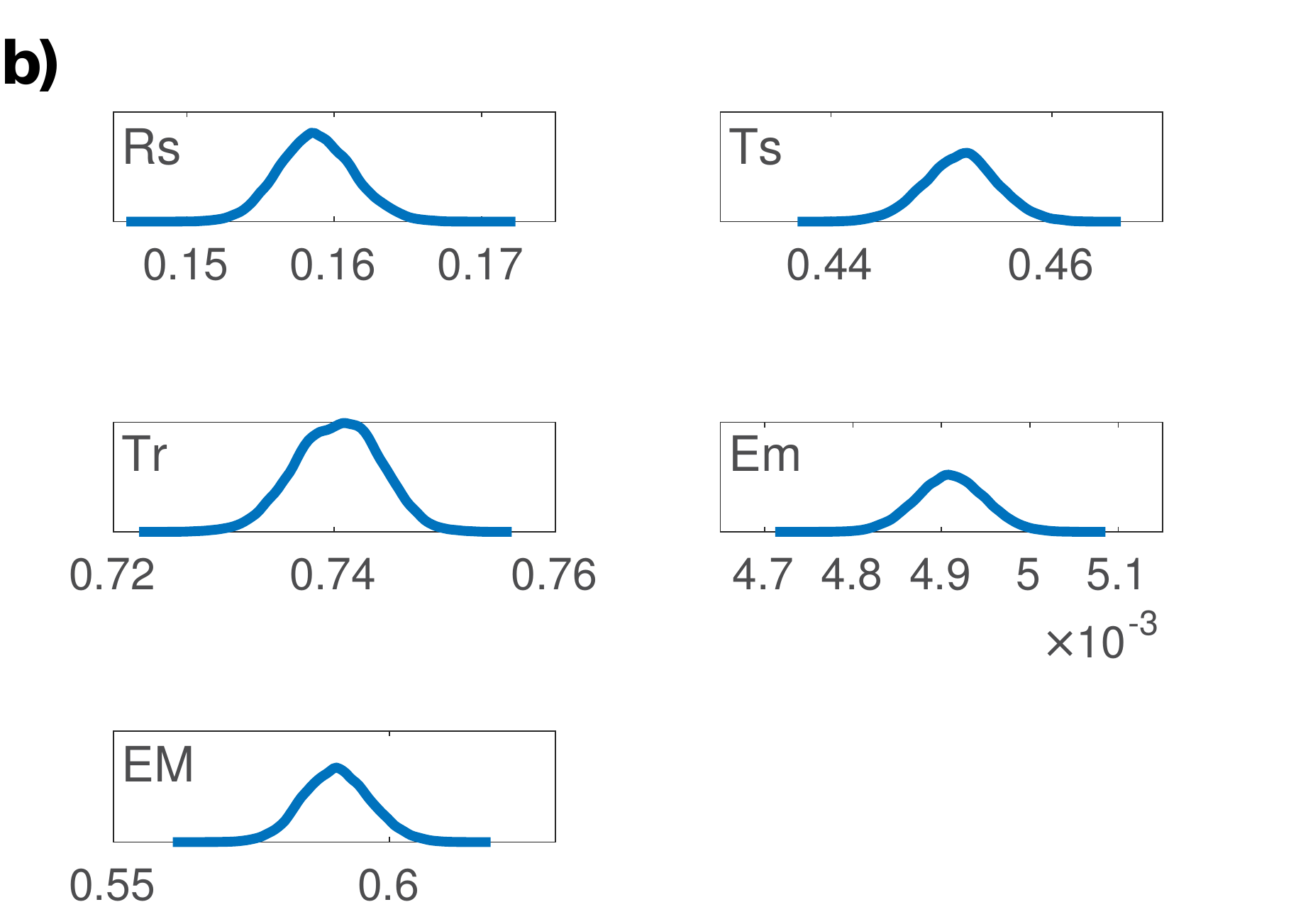}}\vspace{-0.125cm}
\hspace{.6cm}\includegraphics[height = 6cm]{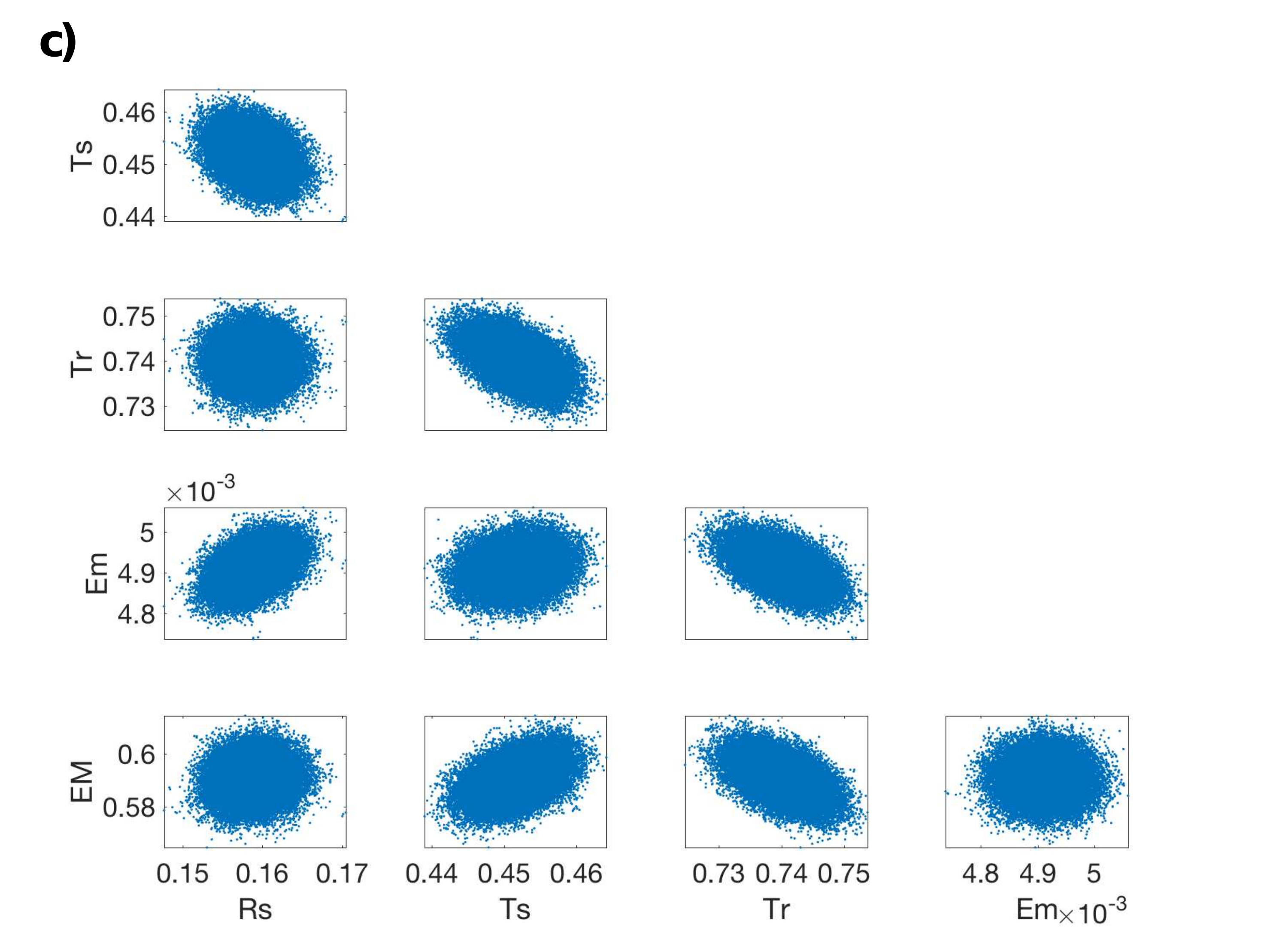}
\caption{MCMC diagnostics from DRAM for Rat 1. {\bf a}) Parameter chains of subset (\ref{eq:subset}) using DRAM with 100,000 sample points.  The first 10,000 samples are considered as the burn-in period and are not used in computing the resulting parameter densities or Bayesian UQ. {\bf b}) Parameter densities of subset (\ref{eq:subset}). {\bf c}) Pairwise correlations of the resulting parameter densities from DRAM.  Each plot can be interpreted as a marginal density by integrating over all of the other parameters not contained in the respective plot. Results for Rats 2 and 3 are similar}
\label{fig:DRAM}
\end{figure}

\begin{figure}[h!]
\centering
\includegraphics[height = 3.4cm]{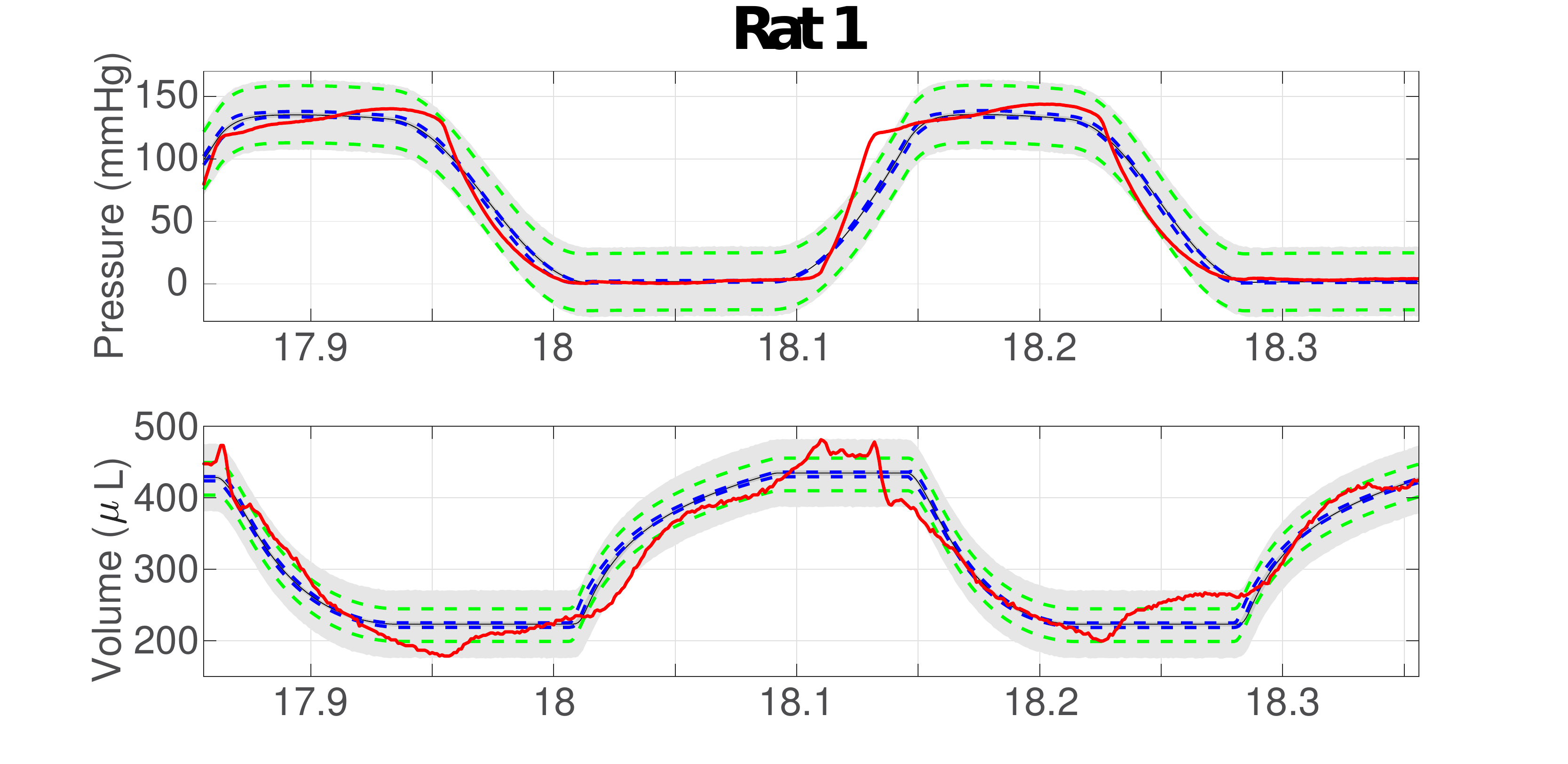}\\
\includegraphics[height = 3.4cm]{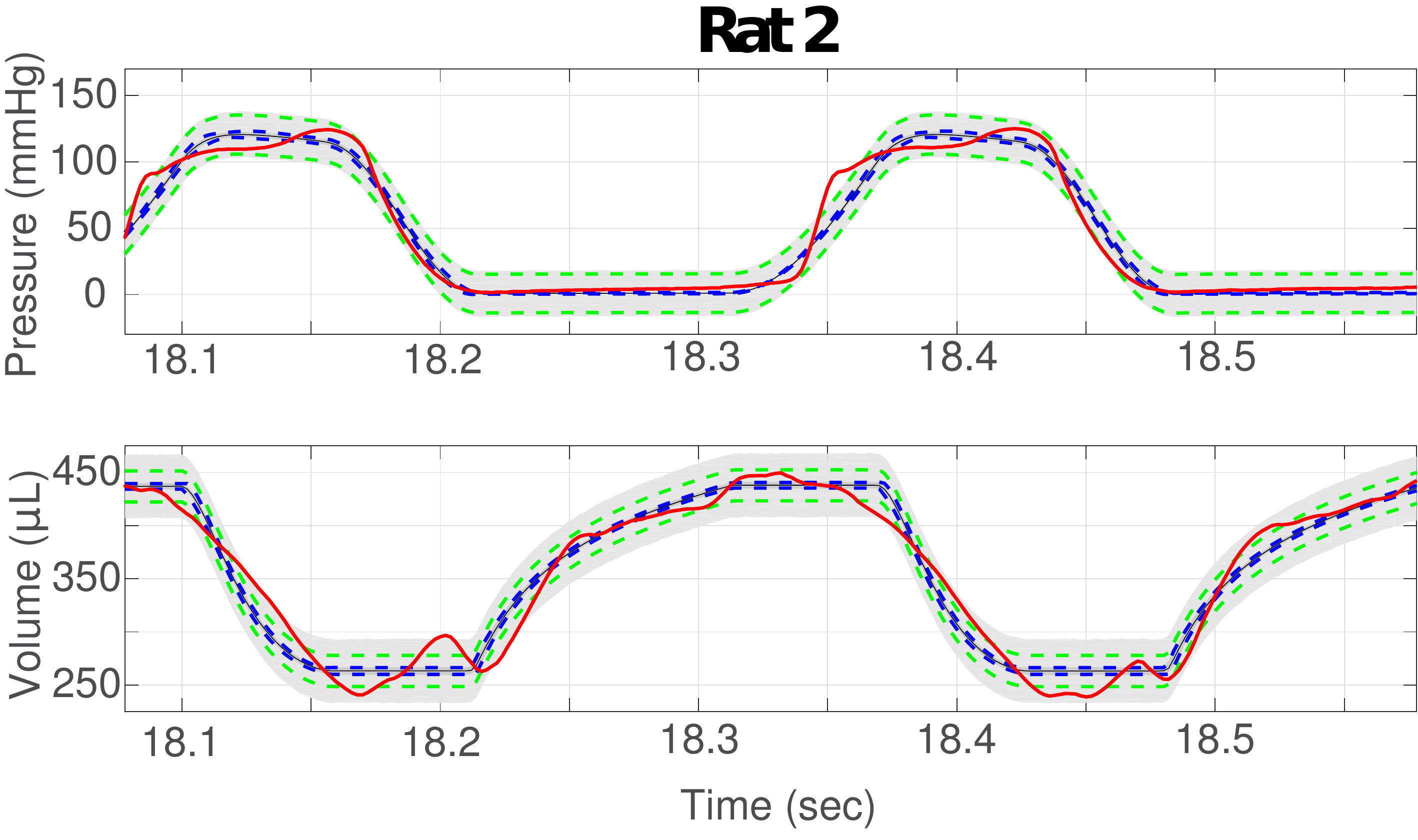}
\includegraphics[height = 3.4cm]{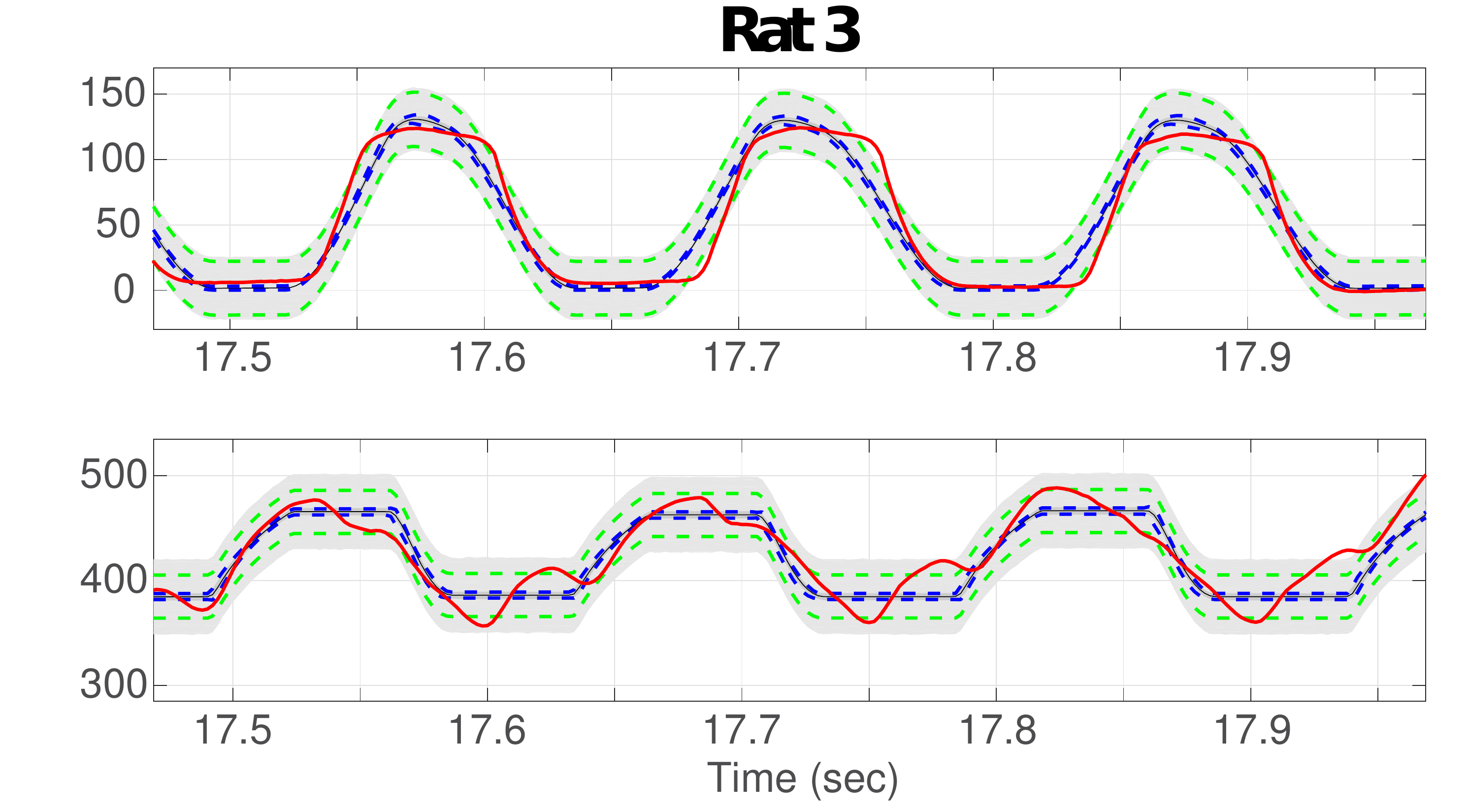}\vspace{-0.4cm}
\caption{Uncertainty propagation results for all three rats.  The red curve is the data, the blue dotted lines show the frequentist confidence interval (\ref{eq:conf}), and the green dashed lines show the frequentist prediction interval (\ref{eq:freq}). The black line is the model evaluated with the means of the DRAM estimated parameter densities, the dark grey band (too narrow to be seen in this figure) is the Bayesian credible interval, and the light grey band is the Bayesian prediction interval.} \label{fig:UQ}
\end{figure}

To perform UQ, we considered both the frequentist formulas stated in Section~\ref{Subsubsection:Freq} and the Bayesian inference using DRAM as described in Section~\ref{Subsubsection:DRAM}.  Using the optimized values, we  applied equations (\ref{eq:conf}) and (\ref{eq:freq}) to compute frequentist confidence and prediction intervals, respectively.  The optimized point estimates were used to construct the prior distribution of parameters as diffuse Gaussian distributions (uninformed prior centered around the optimized point estimate) for the DRAM algorithm. This choice of a prior is justified by the fact there is no information we could use to construct a  more informative prior distribution. Chains of 100,000 sample points were generated using DRAM.  Figure~\ref{fig:DRAM} shows the resulting DRAM-estimated parameter chains, parameter densities, and pairwise parameter correlations for rat 1, similar results were obtained for the other animals.  Frequentist UQ intervals were computed from equations (\ref{eq:conf}) and (\ref{eq:freq}), and Bayesian UQ\ intervals were constructed by solving the model over randomly sampled values from the posterior parameter chains. Figure~\ref{fig:UQ} compares the resulting frequentist and Bayesian UQ for all three animals.

\begin{sidewaystable}[h!]
\caption{Model parameter values for each rat. For each entry (a,b,c) corresponds to Rat 1, 2, and 3 respectively. Bolded parameters are those included in subset (\ref{eq:subset}) and estimated using Levenberg-Marquardt optimization and DRAM.  Reported DRAM values are the means of the resulting parameter sample chains with the 10,000 sample burn-in removed.}
\centering
\begin{tabular}{|l|l|l|l|l|}
\hline 
Parameter & Units   & Nominal $(\cdot10^{-1})$ & Lev-Mar $(\cdot10^{-1})$& DRAM Mean $(\cdot10^{-1})$ \\
\hline
\hline
$R_\text{a}$       & s mmHg/$\mu$l & 0.011, 0.014, 0.012 & -      & -\\
\hline
$\bf{R_\text{s}}$  & s mmHg/$\mu$l & 0.942, 1.155, 1.003 & 1.600, 1.630, 2.196  & 1.588, 1.631, 2.187 \\
\hline
$R_\text{v}$       & s mmHg/$\mu$l & 0.002, 0.001, 0.002 & -      & -\\
\hline
$E_\text{ao}$      & $\mu$l/mmHg   & 9.641, 8.446, 8.535 & -      & -\\
\hline
$E_\text{sa}$      & $\mu$l/mmHg   & 1.193, 1.045, 1.056 & -      & -\\
\hline
$E_\text{sv}$      & $\mu$l/mmHg   & 0.006, 0.003, 0.006 & -      & -\\
\hline
$E_\text{vc}$      & $\mu$l/mmHg   & 0.050, 0.027, 0.049 & -      & -\\
\hline
$\bf{T_{S}}$       & s             & 2.500, 2.500, 2.500 & 4.544, 4.161, 4.166 & 4.518, 4.163, 4.213\\
\hline
$\bf{T_{R}}$       & s             & 6.000, 6.000, 6.000 & 7.309, 6.409, 8.149  & 7.403, 6.409, 8.161 \\
\hline
$\bf{E_{m}}$       & mmHg/$\mu$l   & 0.043, 0.026, 0.042 & 0.050, 0.025, 0.039  & 0.049, 0.025, 0.039 \\
\hline
$\bf{E_{M}}$       & mmHg/$\mu$l   & 5.476, 3.864, 2.595 & 5.965, 4.366, 3.252  & 5.904, 4.364, 3.254 \\
\hline
\end{tabular}
\label{tab:optpar}
\end{sidewaystable}
\clearpage

\section{Discussion}

It is well known that blood vessels dilate and constrict to maintain homeostatic levels of blood flow and pressure. However it is challenging to quantify the functionality of the heart and vasculature. Cardiovascular quantities including volumes, flows, and pressures (e.g. ventricular volume and pressure - studied here) can be measured experimentally, yet the system properties that underlies specific outputs can typically not be measured directly, e.g. peripheral vascular resistance, vessel compliance, or cardiac contractility.  The simple cardiovascular model presented here can provide experimental biologists a framework to estimate these quantities~\cite{young,gelman}.  Using the rigorous approach presented here, we are able to assess model uncertainty that would be essential for clinical application. 

In summary, our step-by-step approach showed that by removing parameter interactions by means of local sensitivity and structured correlation analysis, we can obtain the same subset of practically identifiable model parameters for all three rats. We used nonlinear least squares optimization to estimate this subset of parameters to provide a high quality fit to the data. Finally, we used both frequentist and Bayesian UQ methods to predict  confidence, credibility and prediction intervals around the model output. The approached discussed here was demonstrated on a simple cardiovascular model, yet it is applicable to any model with data that can be formulated as a system of ODEs.

\paragraph{\bf Model Formulation and nominal parameter values}

The model type used in this study is not new, and even though previous studies, e.g.~\cite{Smith04,Lu01,Puelz17,Guyton72} involve more complex interactions (e.g. nonlinear venous compliance, systemic and pulmonary sub-circuits, inductors, etc.), the important conclusion is that to obtain reliable parameter estimates it is essential to compute nominal parameters that are physiologically reasonable. This allows one to determine what parameters are identifiable and to assess model uncertainty. The specific model discussed here includes measurements of left ventricular volume and pressure. Other commonly available data include measurements of arterial pressure, central venous pressure, and cardiac output. For most models, including the one presented here, assumptions must be made since not all information is available, e.g. we used literature estimate to determine the unstressed volume fraction. This is a quantity that is important, but as discussed by Spiegel~\cite{Spiegel16}, no good experimental method exists for determining the stressed-to-unstressed ratio. The model presented here was fitted to left ventricular pressure and volume, the advantage being that both arterial and venous pressures and stroke volume can be determined, while aortic pulse pressure had to be estimated. Many other studies, e.g. the study by Williams et al.~\cite{Williams14} rely on non-invasive measurements typically only available at the arterial side. To identify reliable parameters this model can only be done by incorporating assumptions about venous pressure and cardiac output. As discussed in the study by Pironet et al.~\cite{Pironet16} reliable parameter estimates require measurements of both pressures and left ventricular stroke volume. In general, we show how to determine a subset of identifiable parameters that can be estimated by fitting model output to data.

The model and data analyzed here are pulsatile. To accurately fit model to data in the case of this periodic cardiovascular systems model it is essential to align the phase of the periodic driving function with that of the data. The phase of model predictions is determined by  initial conditions. Since for most compartments only either mean and/or minimum or maximum predictions are available it is difficult to set up initial conditions that provide precise alignment of model predictions to data. Mathematically, we want to initialize the model at ``steady state," so that with constant periodic forcing the solution oscillates around a known mean. Depending on time-scale of the model, typically dictated by compartment elastance, computations should be done over a long enough time to allow the system to reach a steady state. Once steady state is achieved, we show how to systematically determine the optimal shift of the data best aligning the model predictions with the data.  To our knowledge this element has not been addressed in previous studies with respect to cardiovascular models.

Finally, the pulsatile nature of this model is facilitated by periodic forcing of the system combined with valves. This model uses diodes to predict flow into and out of the valves, while this is the most typical formulation~\cite{Pironet16, sun1995mathematical}, the discrete opening and closing of valves make the system of differential equations stiff. While the diodes approach is easier to formulate, it introduces a discontinuity in the system of equations not characteristic of the pressure-dependent resistances~\cite{Williams14} or valves accounting for inertia~\cite{Shi11, smith2004minimal, Mynard12}.  Even when accounting for inertia, the system of ODEs arising in pulsatile cardiovascular models is stiff and requires careful attention when solving numerically; for more details on solving stiff differential equations, see, e.g.~\cite{LeVeque,Iserles}.

Below we discuss results of each component of our step-by-step approach engaged to estimate identifiable parameters and determine model uncertainty.

\paragraph{\bf Sensitivity analysis and subset selection}
 
Most cardiovascular models, as the one formulated here, are over-parameterized compared to available data. While data available for parameter estimation often is dense in time,  only a few states are measured, e.g. the model presented here are based on predictions of left ventricular flow and pressure. Therefore, care must be taken both in the model building phase, by not including more compartments than can be justified, and in the parameter estimation phase, only estimating identifiable parameters. The first step in the analysis phase involves sensitivity analysis. There are many methodologies to consider, and care must be taken since the system of equations often is nonlinear and stiff. For this study we elected to calculate sensitivities using a forward difference methodology, that we validated by reducing our ODE\ solver tolerance and observing that the results converged to stable values. While this approach is easy,  more effort could be put into study the system globally, e.g. using Sobol indices or Morris screening~\cite{ChristianPhD,Eck16}. Alternatively, local results could be obtained by solving sensitivity equations in the complex plane~\cite{martins2003complex} - however this method requires that model be analytic and the diode valve formulation we employ violates this condition.

Parameters were ranked from most to least sensitive using a two-norm averaging over the part of the data used for model validation (i.e. after steady oscillations have been achieved, when the effects of initial conditions disappear). Given the periodic nature of the model studied here this approach makes sense, yet for other model types, generalized sensitivity~\cite{Banks10} may be more appropriate as they also provide information about what part of the data specific parameters influence. 

Results of our sensitivity analysis revealed that $R_{\text{V}}, E_{\text{sv}}, E_{\text{la}},$ and $R_{\text{A}}$ were insensitive compared to the other parameters, as a result we chose to keep those fixed at their nominal values. Another use of sensitivities is to detect if specific model components can be eliminated or should be described in more detail. It should be noted that the sensitives we use reflect how an individual parameter influences the proximity of the model solution to the data rather than providing a metric of how it influences  the model behavior.  

Parameters being sensitive do not equate to being identifiable~\cite{Ottesen13,Miao11}. Thus it is important to combine sensitivity analysis with identifiability analysis, or subset selection (the term used here) to identify relationships among parameters. However, determining if  a specific parameter interaction is identifiable is nontrivial. Pironet et al.~\cite{Pironet16} illustrated that  elastance and resistance parameters form a structurally identifiable relationship given left ventricular stroke volume and pressure data for every compartment in their model. Their model uses a simplified elastance driving function with no parameters and assumed that stroke volume and all compartment pressures were available. Consequently results from the Pironet model cannot directly be translated to our model. Mahdi et al.~\cite{mahdi2014structural} provided some guidelines for constructing arbitrarily complex structurally identifiable spring-dashpot networks, however their results were constrained to linear viscoelastic formulations. Our methodology is motivated by analyzing the practically identifiable components of cardiovascular models to make predictive inference through the use of uncertainty quantification. 

Intuitively one would expect that  parameter interactions hinder the identifiability of a model. However it is crucial to understand, even though intimately related, that correlation and identifiability are distinct concepts. Correlation refers to the precise structure and relationship between interacting parameters, whereas identifiability refers to mapping between the parameter space and the model output, which ideally would be one-to-one. So while minimizing parameter interactions and correlations can be an effective strategy to find an identifiable subset of parameters, it is not an exhaustive or universally applicable strategy for any arbitrary model. We found that approaching the parameter estimation problem from a Bayesian perspective was particularly useful in this regard as it allows one to visualize parameter interactions. It should also be noted that the ideal subset of ``identifiable" parameters  is highly dependent on the type of data available (e.g. pressure in large systemic arteries rather vs.  the left ventricle) and may vary across individuals.

For this study we chose to use structural correlation analysis~\cite{Ottesen13,Miao11}, which is local in nature and it only provides a first order approximation of the parameter correlations, i.e. it is not able  to capture  nonlinear parameter interactions. However, with good initial parameter estimates, this methodology has been shown to work~\cite{ChristianPhD,Ottesen13,Miao11}. In this study, we leverage this limitation using DRAM to estimate the parameters in a Bayesian framework. By representing the parameters as random variables, we can trace the exact shape of the joint densities and can visualize the parameter interactions, as shown in Figure~\ref{fig:DRAM}. Unidentifiable parameter subsets typically take more MCMC iterations to converge and often have more correlation in their pair-wise density plots. We have illustrated that in Figure~\ref{fig:unident} (using a slightly altered subset including $\theta=\{R_\text{s},T_{\text{S}}, T_\text{R}, E_{\text{sa}}, E_\text{m}, E_\text{M}\}$, a subset that structured correlation analysis method determined correlated). Results demonstrate very mild correlations between  $E_\text{M}, E_\text{m}, T_{\text{S}}$ and $T_{\text{R}}$ despite the subset (\ref{eq:subset}) being identifiable. Our recommendation is to combine Bayesian parameter estimation techniques together with asymptotic subset selection methods in an iterative process to refine ones understanding of different model parameterizations and potentially find a practically identifiable subset of parameters.


\paragraph{\bf Optimization}

Similar to numerous other studies, e.g.~\cite{Pope09}, we used a nonlinear least squares methodology to obtain point-estimates for the identifiable parameters. This approach is commonly used for ODE\ models and data describing the  kinetics and/or dynamics of a system~\cite{hovorka2001parameter}. There are many different styles of optimization algorithms with corresponding benefits and limitations.  Gradient-based optimization methods such as Levenberg-Marquart are limited in that they do not search the entire the parameter space while minimizing the cost function. They are designed to find the nearest local minimum relative to the initial parameter estimate, rather than the global minimum. Other optimization routines that do explore the entirety of the parameter space could have been employed (e.g. Nelder-Mead, SPQ algorithm, or interior point optimization), but for well conditioned problems Levenberg-Marquart is computationally very efficient. As a result we put significant effort into {\it a priori} parameterization of the model to yield a well conditioned optimization problem, a topic we have not seen addressed in many cardiovascular models.

\paragraph{\bf Uncertainty Quantification}

Having a high-quality model fit is rarely the end of one's modeling efforts. In particular if the model is to be adapted for analysis of clinical data it is essential to assess the limitations of a model's predictive power. This study used asymptotic and Bayesian methods to determine confidence, prediction and credible intervals. The field of uncertainty quantification is an active area of research~\cite{schiavazzi2016uncertainty, chen2013simulation, sankaran2016uncertainty, pathmanathan2015uncertainty, Paun2018, Eck16} - these references are but a few examples of recent studies applying UQ to models of cardiovascular mechanics.  For the utility that UQ promises, the obvious problem that has been neglected with regard effective implementation is that most methods assume that the mapping from the parameter space to the model output is one-to-one. This is a non-trivial condition that few models satisfy. Furthermore, very few studies have reported results of  identifiability analysis prior to the implementation of the UQ procedure. While one can forcibly propagate the uncertainties of unidentifiable model inputs (parameters and/or data) through a model to obtain uncertainty bounds, the reproducibility and stability of such results are suspect.  

Our analysis, utilizing the estimated subset of identifiable parameters  (\ref{eq:subset}) has allowed us to use both frequentist and Bayesian uncertainty propagation methods to achieve interpretable and tight uncertainty bounds around our model solutions. As a result of using a practically identifiable subset of parameters, our UQ bounds are reproducible and stable across distinct data sets.  Results of analysis showed that asymptotic prediction intervals and the Bayesian prediction intervals are almost identical for pressure data, whereas prediction intervals are slightly wider for the volume data, the latter is likely a result of a higher noise level in the volume data than the pressure data. 

Due to its flexibility in representing parameters as random variables, an advantage of the Bayesian framework is that UQ is an intrinsic feature. Assuming the model is  identifiable  and that  the parameter chains has  converged, credible and prediction intervals can be obtained by extracting model evaluations from the parameter chains. These uncertainty bounds are not subject to any potentially limiting assumption about their distribution, as their frequentist analogs are. The largest limitation to MCMC-type Bayesian approaches is the sheer computational cost. MCMC methods are robust, yet they require thousands of model evaluations to construct a converged posterior parameter distribution. For sufficiently complex models, MCMC may be an impractical route.  As an alternative, sequential Bayesian methods such as particle filtering~\cite{Kaipio2005,LiuWest2001,Pitt1999,Arnold2013} or ensemble Kalman filtering~\cite{Evensen1994,Burgers1998,Evensen2009,Arnold2014} may reduce computational time by evaluating the model from one data point to the next, as opposed to integrating over the entire data set at once.  Another option is to use model emulation within the DRAM evaluation~\cite{rasmussen2006gaussian}. For the cardiovascular model presented here, we can see in Figure~\ref{fig:UQ} that there is very little appreciable difference between the employed statistical approaches. Thus for future studies using this model and corresponding data sets, frequentist UQ alone may be sufficient.  

In conclusion, we developed a model and estimated subject specific nominal parameters using available data and literature. We used local sensitivity analysis and subset selection to determine a set of identifiable parameters that were optimized against left ventricular pressure and volume data from three rats. Subsequently we used asymptotic and Bayesian analysis to estimate prediction and credible intervals, which since done in a close neighborhood of the optimal values were comparable. Given the computational cost associated with Bayesian methods, we propose to identify possible subsets and conduct estimations using both local and Bayesian approaches for a few test animals and then carry out larger population studies using only local methods.

\section*{Appendix}

\subsection*{Model Equations}

The complete system of differential equations describing the rates of change of the compartments of the model analyzed in this study are given as follows:
\begin{eqnarray*}
 \frac{dV_{\text{lv}}}{dt} &=& q_{\text{mv}} - q_{\text{av}} \\[1em]
 \frac{dV_{\text{ao}}}{dt} &=&  q_{av} - \frac{E_{\text{ao}}V_{\text{ao}} - E_{\text{as}}V_{\text{as}}}{R_{\text{A}}} \\[1em]
 \frac{dV_{\text{sa}}}{dt} &=& \frac{E_{\text{ao}}V_{\text{ao}}-E_{\text{sa}}V_{\text{sa}}}{R_{\text{A}}} - \frac{E_{\text{sa}}V_{sa} - E_{sv}V_{sv}}{R_{\text{S}}} \\[1em]
 \frac{dV_{\text{vs}}}{dt} &=& \frac{E_{\text{sa}}V_{\text{sa}}-E_{\text{sv}}V_{\text{sv}}}{R_{\text{S}}} - \frac{E_{\text{sv}}V_{\text{sv}} - E_{\text{vc}}V_{\text{vc}}}{R_{\text{V}}} \\[1em]
 \frac{dV_{\text{vc}}}{dt} &=& \frac{E_{\text{sv}}p_{\text{sv}} - E_{\text{vc}}V_{\text{vc}}}{R_{\text{V}}} - q_{\text{mv}} 
\end{eqnarray*}
where
\[
   q_{\text{av}} = \left\{ \begin{array}{ll} 
       \displaystyle \frac{p_{\text{lv}} - E_{\text{ao}}V_{\text{ao}}}{R_{\text{av}}}  &  \mbox{if valve open, e.g.}  p_{\text{lv}} > p_{\text{ao}}  \\
       \displaystyle 0                                             & \mbox{otherwise (valve closed),}
   \end{array} \right.
\]
\[
   q_{\text{mv}} = \left\{ \begin{array}{ll} 
       \displaystyle \frac{E_{\text{vc}}V_{\text{vc}} - p_{\text{lv}}}{R_{\text{mv}}}  &  \mbox{if valve open, e.g.}  p_{\text{vc}} > p_{\text{lv}}  \\
       \displaystyle 0                                              & \mbox{otherwise (valve closed).}
   \end{array} \right.
\]
and
\[
  p_{\text{lv}} = E_{\text{lv}}(t) V_{\text{lv}},
\]
\[
  E_{\text{lv}}(t) = \left\{ \begin{array}{ll}
       \displaystyle   E_\text{m} + \frac{E_\text{M} - E_\text{m}}{2} \left(1-\cos (\pi t/T_{\text{S}})\right) &  0 < t < T_{\text{S}}  \vspace{4mm}\\
       \displaystyle   E_\text{m} +\frac{E_\text{M} - E_\text{m}}{2} \cos\left(\pi(t-T_{\text{S}} )/(T_{\text{R}}-T_{\text{S}})\right) &T_{\text{S}}<t<T_{\text{R}}, \\
       \displaystyle   E_\text{m}     & T_{\text{R}} < t < T 
       \end{array} \right.
\]

\subsection*{Unidentifiable subset}

Representing parameters as probability distributions can provide information about the identifiability of a multi-dimensional parameter distribution. Plots of parameter chains and densities such as those seen in Figures \ref{fig:DRAM} are the fundamental diagnostic tools  to asses if the MCMC parameter optimizer has converged. The posterior parameter chain plot should ideally be white noise of the distribution, and the posterior parameter density should have a single clearly defined mean. Unidentifiable parameter distributions often have multi-modal distributions that can be observed from plots of the parameter chain and the density. One can interpret a multi-modal distribution as a parameter chain that has not converged, or an unidentifiable parameter. Should the first scenario be the case be, one can run the MCMC algorithm for more iterations until the distribution converges to the true posterior. However, if the converged parameter distribution is still multi-modal, it indicates that multiple values of a parameter can be used to produce the same model output.  By definition, this means that the parameter is unidentifiable. 

\begin{figure}[h!]
\includegraphics[height= 5.3cm]{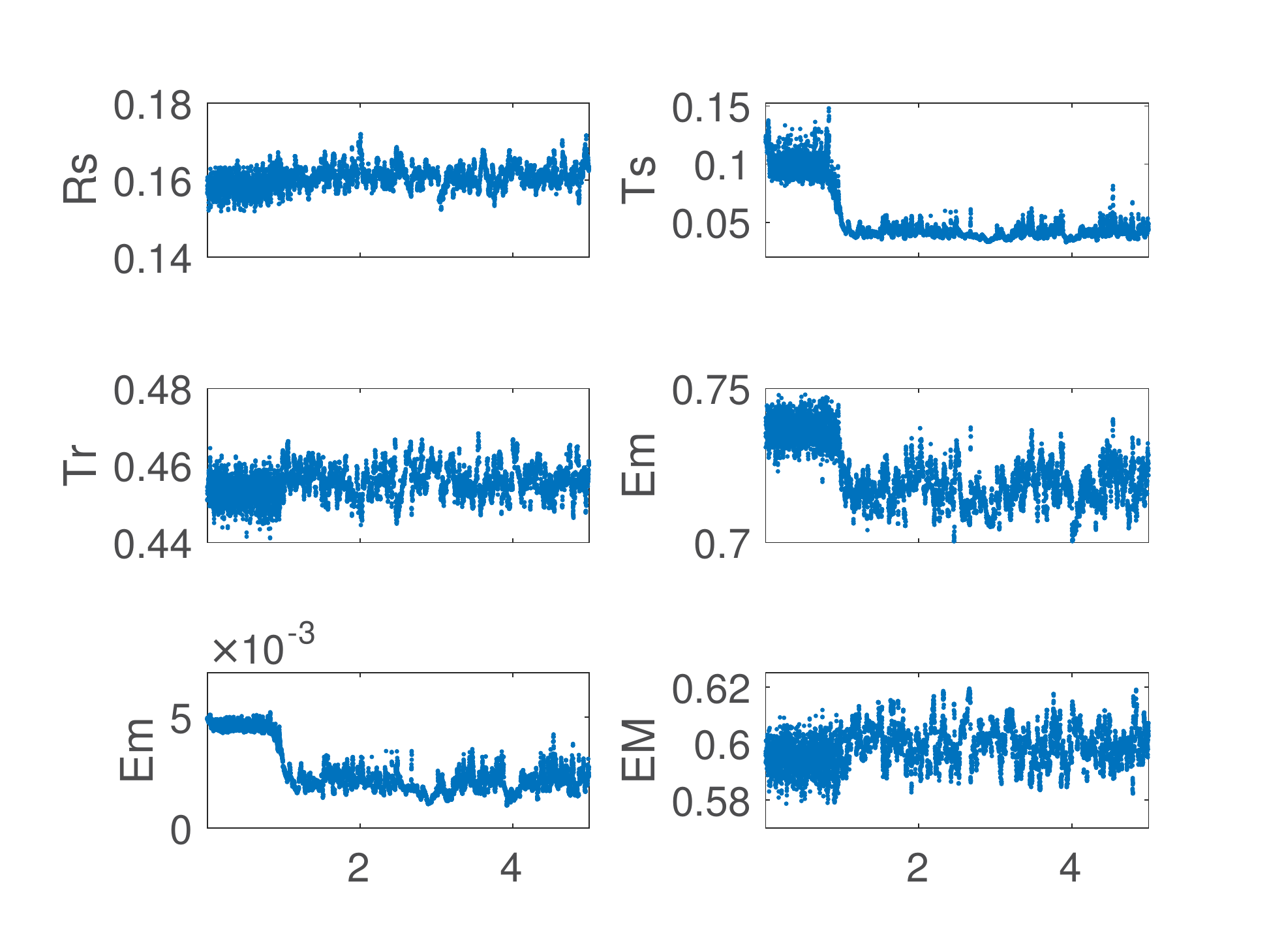}\hspace{-0.5cm}
\includegraphics[height= 5cm]{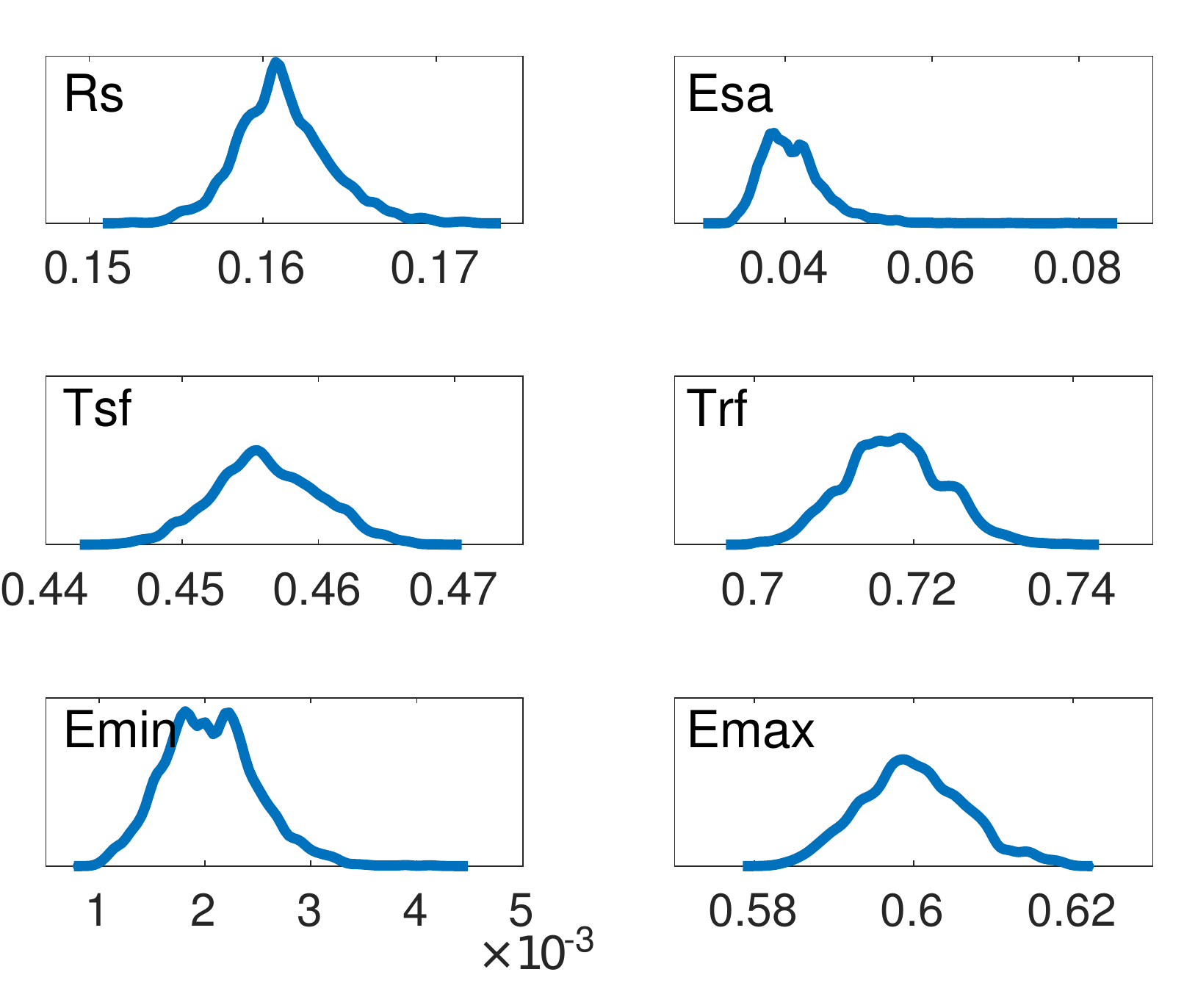}\vspace{-0.125cm}\\
\centerline{\hspace{-1.25cm}\includegraphics[height = 7cm]{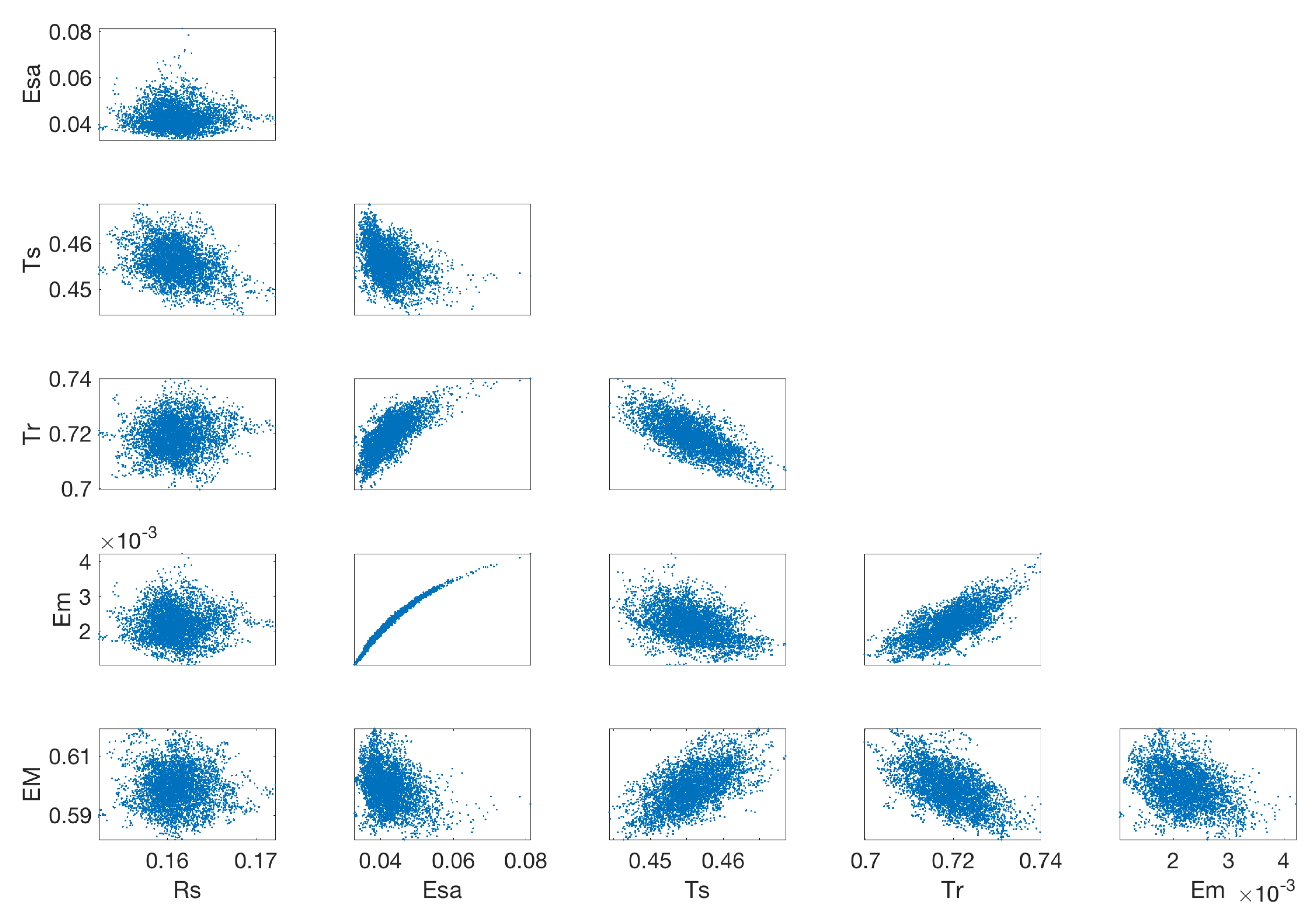}}
\caption{Parameter densities of an unidentifiable subset. Posterior parameter chains were computed using 100,000 iterations of DRAM, with the first 10,000 iterations removed as burn-in.}
\label{fig:unident}
\end{figure}

To illustrate how an unidentifiable subset of parameters can result in multimodal posterior parameter distribution we ran DRAM using a subset which included $E_\text{sa}$. Results from this simulation, shown in Figure~\ref{fig:unident}, revealed that it takes approximately 12,000 iterations for the parameter chains to burn-in (compared with the identifiable subset that  start near the converged distribution), even with this long burn-in the chains are not well mixed. Poorly mixed parameter chains are indicative of an ill-conditioned optimization problem ~\cite{cowles1996markov}. Further note that the parameter densities have lumpy shapes, e.g. parameters $E_\text{sa}$ and $E_\text{m}$ exhibit bimodality. Furthermore, we observe that $E_\text{sa}$ and $E_\text{m}$ have a highly correlated joint density. One may be able to run the MCMC routine for more iterations to get better mixed chains and smooth out the densities, but for an arbitrarily complex model it may not be worth the effort given the raw computational time and resources MCMC routines require. We note that while the modality of posterior parameter distributions can be used to assess the identifiability of a subset of parameters, an individual parameter exhibiting multi-modal behavior is not itself unidentifiable. From these results alone, we could not explicitly determine if $E_\text{sa}$ or $E_\text{m}$ are unidentifiable. Additionally, there are situations where a third parameter with a relatively smooth density may in fact be the unidentifiable culprit. The posterior parameter density is a joint-distribution of all the parameters, so any multi-modal behavior must be considered in the context of the other parameters in the distribution.
\clearpage

\section*{References}

\bibliography{ratbib}

\end{document}